\documentclass[twocolumn]{aa}
\usepackage{graphicx}
\begin{document}
   \title{S$^4$N: A Spectroscopic Survey of Stars in the Solar Neighborhood
   \thanks{Based
 on observations made with the 2.7m telescope at the
   McDonald Observatory of the University of Texas at Austin (Texas), 
   and the  1.52m telescope at the European Southern
Observatory (La Silla, Chile) under the agreement with the CNPq/Observatorio
Nacional (Brazil).}
 \thanks{Tables 5, 6 and 7 are only available in electronic form at 
the CDS via anonymous ftp to cdsarc.u-strasbg.fr (130.79.125.5) or
via http://cdsweb.u-strasbg.fr/Abstract.html}
   }

   \subtitle{The Nearest 15 pc}

   \author{Carlos Allende Prieto
          \inst{1},
          Paul S. Barklem
          \inst{2},
          David L. Lambert
          \inst{1}
          \and
          Katia Cunha\inst{3}
          }

   \offprints{C. Allende Prieto}

   \institute{McDonald Observatory and Department of Astronomy,
              University of Texas, Austin, TX 78712-1083, USA\\
              \email{callende,dll@astro.as.utexas.edu}
         \and
             Uppsala Astronomical Observatory, Box 515, SE 751-20 Uppsala, Sweden\\
             \email{barklem@astro.uu.se}
             %\thanks{Based on observations made at the McDonald and  
             % La Silla (ESO) Observatories.}
         \and
              Observat\'orio Nacional, Rua General José Cristino 77 CEP 20921-400, 
              Rio de Janeiro, Brazil \\
              \email{katia@on.br}
             }

   \date{}

   \abstract{

We report the results of 
a high-resolution spectroscopic survey of all the stars more
luminous than $M_V = 6.5$ mag within 14.5 pc from the Sun. 
The {\it Hipparcos}    
catalog's completeness limits guarantee that our survey is comprehensive and 
 free from some of the selection effects in other samples of nearby 
stars. 
The resulting spectroscopic database, which we have made publicly available, 
includes spectra for 118 stars obtained with 
a resolving power of $R \simeq 50,000$, continuous spectral coverage 
between $\sim 362-921$ nm, and typical signal-to-noise ratios in the range 
$150-600$. We derive
stellar parameters and perform
a preliminary abundance and kinematic analysis of the F-G-K stars 
in the sample. The inferred metallicity ([Fe/H]) distribution is  centered
at about $-0.1$ dex, and shows a standard deviation of $0.2$ dex. 
A comparison with larger samples of {\it Hipparcos} stars, some of which
have been part of previous abundance studies, suggests that our
limited sample is representative of a larger volume of the local thin disk.
We identify a number of metal-rich K-type stars which appear to be very old,
confirming the claims for the existence of such stars in the solar
neighborhood. 
With atmospheric effective temperatures and gravities   
derived independently of the spectra, we find that our classical LTE 
model-atmosphere analysis of metal-rich  (and mainly K-type) stars 
 provides discrepant abundances from neutral and ionized lines of 
several metals. 
This ionization imbalance could be a sign of departures from LTE 
or inhomogeneous structure, which are ignored in the interpretation of the
spectra. Alternatively, but seemingly unlikely, 
the mismatch could be explained by systematic errors 
in the scale of effective temperatures.
Based on transitions of majority species, we discuss abundances of 16
chemical elements. In agreement with earlier studies 
we find that the abundance ratios to iron
of Si, Sc, Ti, Co, and Zn  become smaller as the iron abundance increases
until approaching the solar values, but the trends reverse for higher
iron abundances. 
At any given metallicity, 
stars with a {\it low} galactic rotational velocity 
tend to have high 
abundances of Mg, Si, Ca, Sc, Ti, Co, Zn, and Eu, but low
abundances of Ba, Ce, and Nd.
The Sun appears deficient by roughly 0.1 dex 
in O, Si, Ca, Sc, Ti, Y, Ce, 
Nd, and Eu, compared to its immediate neighbors 
with similar iron abundances.

   }

   \maketitle
%
%________________________________________________________________

\section{Introduction}

The study of the stellar population of the solar neighborhood is of
 interest for  a wide range of investigations.  Nearby stars are  the
only ones for which we can obtain crucial measurements,  such as
parallaxes or angular diameters. They are  
 typically the best candidates for very precise spectroscopic followups,
  such  as searches
for lower-mass companions (stellar,  brown dwarf, or planetary),
circumstellar disks, or solar-like cycles. Their proximity involves obvious 
advantages that should be exploited.

 From a wider perspective, the stellar population of the solar neighborhood 
 may be representative  of larger parts of the Galactic disk.   
 Early studies revealed that the  simplest
models of Galactic chemical evolution
could not match the observed metallicity distribution -- the
so-called G-dwarf problem (van den Bergh 1962). Recent 
investigations have
shown that more complex scenarios, such as those including inhomogeneous
evolution (Malinie et al. 1993) or infall (Chiappini et al. 1997), 
more closely resemble the observations. 
However, almost all the derived metallicity
distributions of samples approximately limited in volume, have
been obtained from photometry (Twarog 1980;
Rocha-Pinto \& Maciel 1996, 1998a; Flynn \& Morell 1997; 
Kotoneva et al. 2002). 
This is worrying, as it has been recognized that some of the 
employed photometric 
indices may be seriously affected by chromospheric 
activity (Rocha-Pinto \& Maciel 1998b; Favata et al. 1997a) and other
systematic effects (Haywood 2002; Reid 2002; 
Twarog, Anthony-Twarog \& Tanner 2002).
These uncertainties can be avoided by using
spectra.  Favata et al. (1997b)  carried out the only existing   
spectroscopic study of a volume-limited sample of nearby stars, 
and found a metallicity distribution for K
dwarfs which is remarkably more metal-rich than those obtained through 
 photometric analyses. 
%Their result
%has been very controversial, and it has been suggested
% that the discrepancy might be traceable to 
% poor statistics (Rocha-Pinto \& Maciel 1998a). 

\begin{figure}[ht!]
\centering
{\includegraphics[width=5.5cm,angle=90]{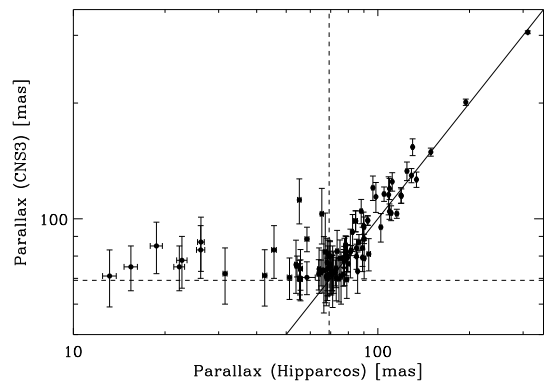}} 
\caption{Comparison of the parallaxes larger than 69.18 mas 
included in  CNS3, against {\it Hipparcos}. {\it Hipparcos} has  revealed  
that a significant fraction of these stars are indeed further 
away from the Sun. The solid straight line has a slope of one.}
\label{cns3}
\end{figure}

We have conducted 
a complete $M_V$-limited volume-limited survey of nearby stars in an attempt to
settle these and other issues.  
In addition to the Catalogue of Nearby
Stars (Gliese \& Jahrei\ss\ 1991, hereafter CNS3) -- used by most of the previous 
studies-- we have taken advantage of the  {\it Hipparcos} catalog (ESA 1997). 
  {\it Hipparcos}
is complete down to $V=7.3$\footnote{This is a lower limit; the actual
values depend on galactic latitude and spectral type.}.  
Restricting the study to stars 
brighter than $M_V = 6.5$ ($\simeq$ K2V), {\it Hipparcos}' completeness 
reaches out to 14.5 pc. 
Figure \ref{cns3} shows 
that  nearly 25 \%  
of the stars included in CNS3 
within 14.5 pc from the Sun (parallaxes larger than 69.18 mas) are 
in fact farther away according to the  
 accurate parallaxes measured by {\it Hipparcos}. 
 
\begin{figure}[ht!]
\centering
{\includegraphics[width=5.5cm,angle=90]{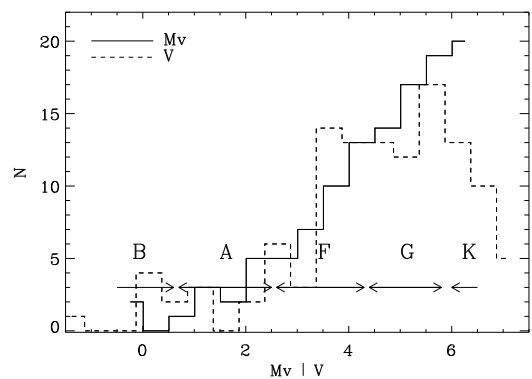}}
\caption{Distribution of absolute ($M_V$; solid line) and 
visual ($V$; dashed line) magnitudes of the sample. Approximate 
spectral types for dwarfs are assigned to the different ranges of $M_V$ 
as reference.}
\label{dist}
\end{figure}

High-dispersion high signal-to-noise
optical spectra were acquired for 118 stars: all the
 objects  more luminous than $M_V=6.5$  
  within 14.5 pc from the Sun. 
Figure \ref{dist} shows the sample's distribution of absolute and 
visual $V$ magnitudes:
  $M_V$ (solid line) and $V$ (dashed line). As a
reference, spectral types for dwarfs are assigned to  different ranges in $M_V$.
This sample, as a result of the dependence of 
 stellar lifetime on mass and the shape of the initial mass function,
 is dominated by the coolest stars within our range of spectral types.
 This should be borne in mind when comparing to most other high-resolution
 spectroscopic studies of disk stars, which tend to select warmer targets.

We provide a high-quality spectroscopic archive of these 
{\it nearest} stars. The data acquisition, reduction and archival, are
 the subject of Section \ref{observations}. 
Many properties of these stars have been compiled: radial velocities, 
parallaxes, and proper motions. Other properties
have been determined afresh: stellar atmospheric parameters, kinematics, 
ages, and chemical abundances. The 
 resulting catalog is described in \S \ref{catalog}. 
Section \ref{balance} examines the consistency of the inferred abundances.
Section \ref{discussion} discusses our findings. Section \ref{extended}
compares the kinematics and metallicities of the nearest stars with larger
samples of thin-disk stars, and \S \ref{wrapping} presents a short summary
and provides suggestions for future work.

\section{Observations, data processing, and final archive}
\label{observations}

The observations  were obtained with the  Harlan J. Smith 2.7m 
telescope at McDonald Observatory and the ESO 1.52m telescope on La Silla.
The spectra  have a resolving
power of $\sim 5 \times 10^4$, fully cover the optical range, and extend 
into the near IR.  A total of six observing 
campaigns were necessary, as summarized in Table 1.

   \begin{table}
      \caption[]{Observations.}
         \label{runs}
     $$ 
         \begin{array}{p{0.5\linewidth}l}
            \hline
            \noalign{\smallskip}
            Observatory      &  {\mathrm{Dates}}\\
            \noalign{\smallskip}
            \hline
            \noalign{\smallskip}
           La Silla &   {\mathrm{October~ 2000}}     \\            
            McDonald &    {\mathrm{December~ 2000}}  \\
            McDonald &   {\mathrm{April-May~ 2001}}  \\
           La Silla &   {\mathrm{May ~2001 }}  \\            
            McDonald &   {\mathrm{September~ 2001}}  \\   
           La Silla &    {\mathrm{September~ 2001}}  \\                          
            McDonald &    {\mathrm{October ~ 2001}}  \\            
           La Silla &     {\mathrm{November~ 2001}}  \\                 
          \noalign{\smallskip}
            \hline
         \end{array}
     $$ 
%\begin{list}{}{}
%\item[$^{\mathrm{a}}$] This is footnote a
%\end{list}
   \end{table}

The McDonald spectra were obtained with the 2dcoud\'e
spectrograph
(Tull et al. 1995). This cross-dispersed echelle spectrograph provides, 
in a single exposure, full coverage from about 360 nm 
 to $\sim 510$ nm, but has increasingly larger gaps between
redder orders. This problem was circumvented by using two different 
overlapping settings. %, as approximately shown in Fig. \ref{setup}. 
The detector was TK3, 
a thinned $2048\times2048$ Tektronix CCD with 24 $\mu$m pixels, which was 
installed at the  F3 focus. We used 
 grating E2, a 53.67 gr mm$^{-1}$ R2 echelle from Milton Roy Co. and
  slit \#4, which has a central width of 511 $\mu$m (or 
 approx. 1.2 arcsec on the sky).

The ESO (La Silla) data were acquired with FEROS (Kaufer et al. 2000),
a fiber fed cross-dispersed  spectrograph with an R2 79 gr mm$^{-1}$
echelle grating. 
The detector was a $2048\times4096$ EEV CCD with 15 $\mu$m pixels. 
The FEROS resolving power was about 45000, which is roughly 20 \%   
lower than that of the 2dcoud\'e spectra.
This instrument has a fixed configuration, but provides
better coverage than the 2dcoud\'e for a single setting: 
approximately from 350 to 920 nm, leaving two
gaps at $\sim 853$ and 870 nm. 
Most of the observations from the May and September 2001 runs were
obtained in  `object-calibration' mode with FEROS: the Th-Ar lamp was
on during the stellar observations and its light fed
to the spectrograph. This mode was selected due to the fact that the
spectra were obtained in between observations for a different program, for
which an accurate wavelength calibration was crucial.
This turned out to be a disadvantage as, in
some cases, scattered light from Ar lines in the red affected the stellar 
spectra. 

\begin{figure}
\centering
{\includegraphics[width=5.5cm,angle=90]{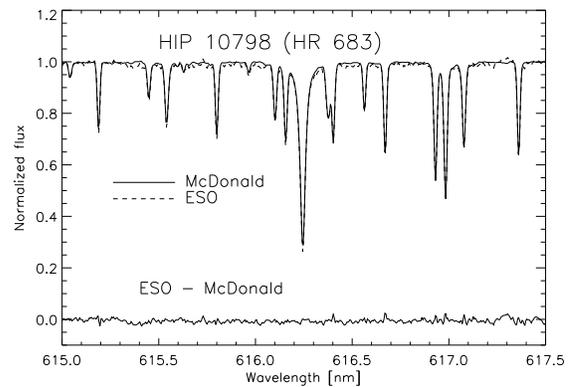}} 
\caption{Continuum corrected spectra of HIP 10798 obtained with the
2dcoud\'e spectrograph on the 2.7m telescope at McDonald Observatory and 
the FEROS spectrograph on the 1.52m telescope at ESO. The McDonald spectrum
has been convolved with a 1.5 pixels FWHM Gaussian to compensate 
for the higher resolution
 of the 2dcoud\'e observations. The difference between the two spectra is
 also shown with a solid line (meandering about zero).}
\label{mcd_eso}
\end{figure}

Aiming for very high signal-to-noise ratios, at least two exposures per setting 
were obtained at McDonald. Due to a more limited observing time, a lower
$S/N$ was reached for the stars observed from ESO, for which a single
exposure was typically obtained. About 2/3 of the sample 
was observed from McDonald. Fig. \ref{mcd_eso} compares the McDonald and 
ESO spectra of HIP 10798 in the region of the Ca I line at 616.2 nm.

%\begin{figure*}
%\centering
%{\includegraphics[width=8.5cm,angle=90]{./setup.ps}} 
%\caption{Spectral setups employed in the McDonald observations. 
%The boxes labeled as RED and BLUE show the size of the Tektronik CCD
% on the focal plane. 
%The horizontal 
%rules indicate the free spectral range for each order. Order numbers 
%and the beginning and ending wavelengths  for some  orders  
%(in \AA) are printed.}
%\label{setup}
%\end{figure*}

The spectra were reduced with 
IRAF\footnote{IRAF is distributed by the National Optical Astronomy Observatories,
    which are operated by the Association of Universities for Research
    in Astronomy, Inc., under cooperative agreement with the National
    Science Foundation.}, using the {\it imred}, {\it 
ccdred}, and {\it echelle} packages. The processing included overscan 
subtraction, flatfielding, scattered light removal, optimal extraction 
and wavelength calibration. Th-Ar observations
were very frequent at McDonald. These were combined 
into an atlas available from the 
WWW\footnote{http://hebe.as.utexas.edu/2dcoude/thar} 
(Allende Prieto 2002a).
The instrumental response, dominated by the blaze function 
characteristic of an echelle spectrograph, 
was carefully removed by two-dimensional
modeling following Barklem et al. (2002).
The {\it blue} and {\it red} setups (McDonald spectra) were combined.
% after correcting for (mainly Earth's) projected velocity variations.
Finally, the orders were merged into a continuous spectrum.
%, where 
%regions that were observed multiple times weightly combined the available
%observations. 

\begin{figure}
\centering
{\includegraphics[width=5.5cm,angle=90]{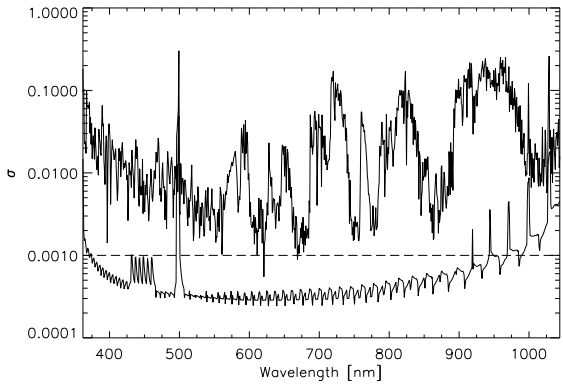}} 
\caption{Contributions to the error in the normalized fluxes for the day light
spectrum acquired with the 2dcoud\'e: continuum placement (thick solid
line; derived by
direct comparison with the solar flux atlas), Poissonian limit 
(thin solid), and
pixel-to-pixel sensitivity corrections (horizontal dashed  
line).}
\label{sigma}
\end{figure}

In an attempt to produce a highly homogeneous archive, the final spectra
were velocity corrected using the central 
wavelengths of hundreds of solar atomic lines as 
reference (Allende Prieto \& Garc\'{\i}a L\'opez 1998). The McDonald 
spectra were then
truncated to the wavelength range 362--1044 nm, while the ESO spectra
were restricted to the range 362--921 nm. Errors in the fluxes were
estimated from three sources: the Poisson noise ($\sqrt{\rm photons}$), 
the continuum placement, and an absolute limit imposed by the best correction
of pixel-to-pixel sensitivity variations we can achieve. The uncertainty in
the continuum placement for McDonald spectra was estimated by comparing 
our `sky-light' spectrum, obtained through a port that brings day light
from the sky to the slit room, 
with the solar flux atlas of Kurucz et al. (1984). For the ESO spectra
 we compared spectra of stars
observed both from ESO and McDonald. Fig. \ref{sigma} shows the different
error contributions adopted for the McDonald sky (solar) spectrum.

In Fig. \ref{sigma}, the largest differences between the McDonald sky
spectrum and the Kurucz et al. (1984) solar atlas (thick solid line) 
are due to 
difficulties in selecting uniquely the position of the pseudo-continuum
in the blue wavelengths due to severe line absorption, 
and to the presence of telluric features and different observing
conditions in the red and infrared part of the spectrum.
The zigzagging shape of the Poissonian limit curve (thin solid line)
is caused by the combination of  multiple observations of the same
spectral segments 
in contiguous orders of the same setting, and also by 
the overlapping wavelength coverage between the two settings.
The wide peak at $\sim$ 500 nm corresponds to the so-called `picket fence'
in the McDonald spectrum -- stray light reflected by the CCD 
back to the grating.  
It is also worthwhile to mention that the ESO spectra at 
and near H$\alpha$ have large error bars due to a reflection 
from the fiber, and thus the H$\alpha$ spectrum in the ESO spectra 
is unreliable.  An error spectrum has been assembled along with 
each stellar spectrum, which reflects the combined error as discussed above.

The spectra are archived in FITS\footnote{Flexible Image Transport System;
see e.g. http://fits.gsfc.nasa.gov/} format, and are available through the
internet at the 
S$^4$N site\footnote{http://hebe.as.utexas.edu/s4n/, 
 http://www.astro.uu.se/$\sim$s4n/}.

\section{The S$^{\rm 4}$N catalog}
\label{catalog}

\subsection{Stellar parameters}
\label{pars}

The `traditional' procedure to derive stellar parameters for a
spectroscopic analysis involves the use of photometry and the spectra
themselves. Among the photometric calibrations available to derive
the stellar effective temperature, those based on the 
infrared flux method (IRFM, Blackwell et al. 1977) 
have been proven very precise
and were adopted here -- always assuming that the reddening must be
negligible within 15 pc. 
Optical and
infrared colors are only weakly sensitive to other stellar parameters.
The extreme accuracy of the parallaxes provided by
the {\it Hipparcos} mission for the nearest stars, can be exploited to
tightly constrain their brightness and gravity. Such 
gravities are practically independent from model atmospheres.
 Our atmospheric stellar parameters are,
 therefore, independent from the high-resolution spectra, and can
 be considered fairly robust against the subtleties involved in the 
 interpretation  of the high-resolution spectra.

\subsubsection{Photometry}

Visual magnitudes, $b-y$ and $c_1$ indices were compiled from 
Hauck \& Mermilliod (1998). Sometimes this source does not provide 
uncertainties for the data, and we adopted 0.01 mag.
When $V$ magnitudes were not available from Hauck \& Mermilliod,
we searched Kornilov et al. (1991), correcting their values using

\begin{equation}
V  =   V_{\rm K}    - 0.0019   - 0.061  (B-V)_{\rm K}   + 0.072 (R-I)_{\rm K}, 
\end{equation}

\noindent where the subscript `K' indicates Kornilov et al. 
magnitudes, for which we adopted an error of 0.01 mag. 
This correction 
follows the recommended transformation by 
Kornilov, Mironov, \& Zakharov (1996).
 When a star was not in any of these sources, we used 
$H_p$ from {\it Hipparcos}, the {\it Hipparcos} $(B-V)$ 
and the Simbad  $(U-B)$
\footnote{Except for
HIP 17420 with no $(U-B)$ from Simbad, for which we adopt $+0.49$ as 
for a similar star: HIP 84405.}
with Harmanec's (1998) formulae to estimate $V$, and assign either 
0.01 mag of error or
the $1\sigma$ uncertainty in $Hp$, whatever is larger.
We also compiled  
%$B-V$, $e\_(B-V)$, $V-I$, $e\_(V-I)$, $Hp$, $Scatter\_Hp$,  
%and $V(hip)$
$B-V$, $V-I$, $H_p$, $V$, and their uncertainties 
from the {\it Hipparcos} catalog. 
These data are available on the S$^4$N web site.

\subsubsection{Effective temperature}

We used the Str\"omgrem photometry and the $(B-V)$ index  
with  the Alonso et al. (1996; 1999a)
 calibrations to derive $T_{\rm eff}$. 
The IRFM is based on the comparison between the bolometric flux and
the monochromatic flux at a selected wavelength in the infrared.
%, which
%makes it very robust against the uncertainties that affect the construction of
%model atmospheres. 
Following the application of the IRFM to derive $T_{\rm eff}$ 
for a large number of stars, Alonso et al. derived photometric calibrations.
Use of such calibrations does not require  bolometric fluxes 
and benefits from the statistical elimination of outliers. Our effective
temperatures are based on such photometric calibrations, which 
are also dependent (although weakly) on the stellar metallicity and
surface gravity.
%In the beginning, we do not have estimates of 
%[Fe/H] and $\log g$, 
We begin by assuming [Fe/H]$=0$\footnote{For an element `El' we define 
[El/H] $ = \log \left( \frac{\rm N(El)}{\rm N(H)} \right) - 
\log \left( \frac{\rm N(El)}{\rm N(H)} \right)_{\odot}$, where N represents
number density.}.
Then, other stellar parameters, including [Fe/H], 
are determined from the spectrum (see below), and a second 
iteration is applied to derive the $T_{\rm eff}$s  in Tables 2-4.
The error bars assigned to the temperatures have
been derived from propagating the uncertainties in the photometry
 through Alonso et al.'s polynomials.
We found the calibration based on $(b-y)$ and $c_1$ to provide 
slightly warmer temperatures than the values from $(B-V)$ 
by an average of $42 \pm 6$ K, with an rms difference of 61 K.

An independent assessment of the scale of effective temperatures
can be obtained by exploiting the sensitivity of the wings of Balmer 
lines to the thermal structure of deep atmospheric layers.
Echelle spectrographs are not ideally suited for measuring the shape
of broad features such as the wings of Balmer lines, but as 
we explained in \S \ref{observations}, we paid particular attention to
the continuum normalization of our spectra.   
The H$\alpha$ profiles in the ESO spectra are unreliable, as
suggested  by the difficulties encountered in the normalization process.
However, we deem H$\beta$ for these spectra as `reliable', and both 
H$\alpha$ and H$\beta$ in the 2dcoud\'e spectra, as `highly reliable'. 
We calculated synthetic
profiles for H$\alpha$ and
H$\beta$ as described by Barklem et al. (2002).
Stark broadening is described by the model-microfield method calculations 
of Stehl\'e \& Hutcheon (1999) and self-broadening by calculations of 
 Barklem, Piskunov \& O'Mara (2000a,2000b). 
Radiative broadening is included, and so is an estimate 
of the helium collisional broadening 
(based on rescaling of Barklem \& O'Mara 1997), although both effects are small. 
The model atmospheres  were computed with the same version of the MARCS 
code as the models employed to derive abundances 
(see \S \ref{abu}).  The only differences affect the convection parameters.  
Convection is described by the mixing length treatment, where for our 
purposes the most important parameters are the mixing-length $l$, which 
is usually expressed in units of the pressure scale height as the 
mixing-length parameter $\alpha= l/H_p$, and a second parameter $y$ 
describing the temperature structure within convective elements 
(see Henyey et al. 1965 for details).  The standard set of
 parameters in the MARCS code are $\alpha=1.5$ and $y=3/(4\pi)^2$, and
these are adopted for the models used in the abundance analysis.
As seen in Barklem et al. 
(2002; see also Fuhrmann 1998 and references therein), a better match to 
the observed profiles can be achieved by reducing the mixing length. 
Thus, for our analysis of hydrogen lines we adopted a model grid with
$\alpha=0.5$ and $y=0.5$ as in Barklem et al. (2002).  We note that  
the different values have a negligible impact on the overall 
temperature scale (Barklem et al. 2002).  The effective temperatures
were determined by $\chi^2$ fitting as described in Barklem et al. (2002).

\begin{figure}
\centering
{\includegraphics[width=5.5cm,angle=90]{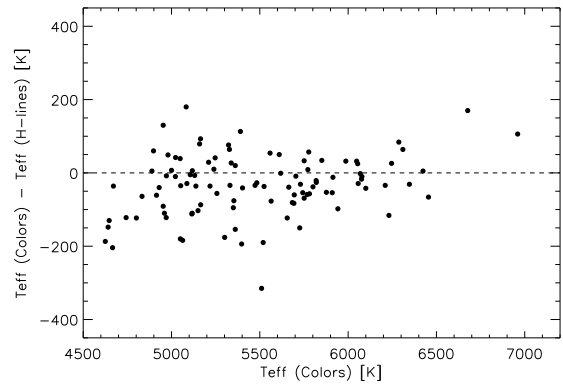}} 
\caption{Differences between the color-based effective temperatures 
derived from IRFM calibrations and those from fitting the wings of
H$\alpha$ and H$\beta$.}
\label{balmer}
\end{figure}

 Figure \ref{balmer} shows the differences between  the IRFM-based
 $T_{\rm eff}$s and those from Balmer lines (weighted average of H$_\alpha$ and
 H$_{\beta}$). The agreement is satisfying; Balmer lines temperatures
 are warmer with a mean difference of
 just 35 $\pm$ 15 K, and an rms difference of 84 K. 
 Fig. \ref{balmer} indicates that the Balmer-line
 temperatures may be slightly higher for stars with $T_{\rm eff}< 4900$ 
 K\footnote{A particular remark should be made about the $T_{\rm eff}$
 for HIP 37826 (Pollux), for which we obtain $4662  \pm  75$ K 
 from $(b-y)$, and  $4684 \pm 155$ from $(B-V)$. Although we adopt the
 weighted average for consistency, both the Balmer profiles and
 the literature, in particular the 
 direct (rather than based on calibrations of $T_{\rm eff}$ vs. colors) 
 value from Alonso et al. (1999b), show a tendency toward a slightly 
 higher value than from $(B-V)$.}.
 A similar effect was also noted by Barklem et al. (2002) 
 for a different sample, but, as they remarked, the reader should bear
 in mind that errors in Balmer-line temperatures are largest in that
 domain due to a) the reduced sensitivity of the line wings, and b) an increasing
 contamination by metallic lines. The second effect can potentially induce
 a bias in the observed sense. Balmer-line $T_{\rm eff}$s might also be
 slightly cooler for the warmer stars in our sample.
 
 Recently, Kovtyukh et al. (2003) have derived effective temperatures for 
 late-type dwarfs using line-depth ratios. From 13 stars in common, we find
 their temperatures systematically higher by $119 \pm 12$ K ($\sigma = 43$ K)
 in the range $5000 < T_{\rm eff} < 6000$ K. An aberrant exception takes
 place for HIP 23311 (HR 1614), for which they derived 4945 K, but our
 preferred value is 4641 K. We note that Feltzing \& Gustafsson (1998) found
 4625 K from Str\"omgrem photometry, and Feltzing \& Gonzalez (2001) 
 arrived at 
 4680 K from the excitation equilibrium of neutral iron and 
 nickel\footnote{Note, however, that these authors 
 determined a gravity for this star based on the iron ionization
 balance which is about 0.6 dex lower than our value (see the discussion in
 \S \ref{balance}).}.
 From fitting the wings of Balmer lines, we obtain an intermediate value of 
 $4789 \pm 110$ K. 
%Other photometric 
% values from the literature are: 4750 K (Oinas 1974), 
%  4846 K (Taylor 1994). 

\subsubsection{Metallicity, surface gravity, age, micro and macroturbulence}
\label{others}

\begin{figure}
\centering
{\includegraphics[width=5.5cm,angle=90]{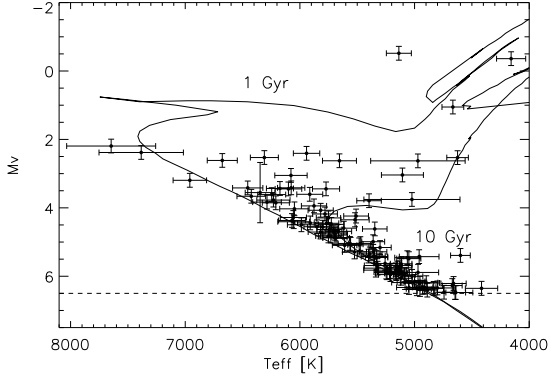}} 
\caption{Distribution of the sample stars in the $M_V$ versus $T_{\rm eff}$
plane.}
\label{parameters}
\end{figure}

Metallicities are first estimated from the automated system described by
Allende Prieto (2003). Gravities (and ages when possible) 
are then estimated from $T_{\rm eff}, M_V$, and [Fe/H], 
by using Bertelli et al. (1994) isochrones
in a similar manner to  Allende Prieto et al. (2003) and
Reddy et al. (2003).
More details are provided in Appendix A.
A first abundance run with the selected spectral line profiles and 
using MARCS models (see \S \ref{abu})
provided refined values for [Fe/H], the microturbulence ($\xi$), and the 
macroturbulence ($\eta$),
that were also used to determine the final $T_{\rm eff}$ (from photometry) 
and gravities (from the isochrones). 
The final stellar parameters are listed in Tables 2-4. 
Fig. \ref{parameters} displays the distribution of the stars 
in the $M_V$ vs. $T_{\rm eff}$ plane. 
Two solar-composition isochrones of ages 1 and 10 Gyr are also shown.

The derived values for the microturbulence ($\xi \equiv \sqrt{2} \sigma$,
where $\sigma$ is the standard deviation of a Gaussian distribution)
are a function of the $T_{\rm eff}$ and $\log g$ of the stars, as
previously found by other authors (e.g. Nissen  1981; 
Edvardsson et al. 1993; Reddy et al. 2003). For our sample, the expression 

\begin{eqnarray}
\begin{tabular}{cc}
$\xi = 1.645$  & $+ ~ 3.854 \times 10^{-4} ~~(T_{\rm eff} - 6387)$  \\
 	& 	 \\
	&       $-0.6400 ~~(\log g - 4.373)$ ~~~~  \\
 	&		\\
 $- 3.427 \times 10^{-4}$ & $(T_{\rm eff} - 6387)~(\log g - 4.373)$ ~km s$^{-1}$
 \\
\end{tabular}
\end{eqnarray}

\noindent predicts $\xi$  
with an rms scatter of 0.14  km s$^{-1}$.

   \begin{table}
      \caption[]{Stars in the sample.}
         \label{runs}
     $$ 
         \begin{array}{llllllll}
            \hline
            \noalign{\smallskip}
            {Hipp. \#}  & T_{\mathrm{eff}} &  \sigma  &  {\mathrm{log}g} & \sigma & {\mathrm{[Fe/H]}^{\mathrm{a}}} & \xi  & \eta \\
            \noalign{\smallskip}
	     & ({\mathrm K}) &    & ({\mathrm cgs}) &	&  
	     & ({\mathrm{km~s}}^{-1}) & ({\mathrm{km~s}}^{-1}) \\            
            \hline
            \noalign{\smallskip}
Sun  & 5777  &   10   &   4.437  &    0.001	& 7.55 &  1.25 & 3.27 \\
171 & 5361 & 119 & 4.610 & 0.040 & 6.78 & 1.21 & 3.01\\
544 & 5353 & 97 & 4.553 & 0.047 & 7.60 & 1.25 & 3.75\\
1599 & 5851 & 114 & 4.468 & 0.067 & 7.30 & 1.28 & 3.68\\
2021 & 5772 & 118 & 3.999 & 0.098 & 7.43 & 1.38 & 4.02\\
3093 & 5117 & 94 & 4.576 & 0.040 & 7.65 & 0.93 & 2.94\\
3765 & 4980 & 92 & 4.652 & 0.051 & 7.24 & 1.17 & 3.18\\
3821 & 5801 & 132 & 4.470 & 0.074 & 7.24 & 1.29 & 3.36\\
4148 & 4898 & 101 & 4.631 & 0.042 & 7.36 & 1.13 & 3.21\\
5336 & 5323 & 119 & 4.665 & 0.043 & 6.74 & 1.34 & 2.82\\
7513 & 6100 & 116 & 4.166 & 0.103 & 7.60 & 1.60 & 6.09\\
7751 & 5083 & 117 & 4.632 & 0.049 & 7.36 & 1.10 & 3.20\\
7918 & 5768 & 107 & 4.422 & 0.079 & 7.49 & 1.22 & 3.50\\
7981 & 5138 & 93 & 4.602 & 0.044 & 7.45 & 1.04 & 2.94\\
8102 & 5328 & 110 & 4.622 & 0.054 & 7.03 & 1.14 & 2.82\\
8362 & 5257 & 98 & 4.576 & 0.048 & 7.50 & 1.06 & 2.90\\
10138 & 5159 & 100 & 4.615 & 0.048 & 7.32 & 1.12 & 3.23\\
10644 & 5664 & 115 & 4.465 & 0.088 & 7.06 & 0.86 & 3.75\\
10798 & 5338 & 105 & 4.632 & 0.053 & 7.08 & 1.12 & 2.81\\
12777 & 6210 & 118 & 4.350 & 0.066 & 7.51 & 1.49 & 5.60\\
12843 & 6231 & 145 & 4.339 & 0.067 & 7.59 & 1.61 & 13.04\\
13402 & 5087 & 96 & 4.598 & 0.040 & 7.65 & 1.33 & 3.77\\
14632 & 5877 & 110 & 4.271 & 0.099 & 7.59 & 1.29 & 3.78\\
14879 & 6078 & 141 & 3.982 & 0.098 & 7.32 & 1.38 & 4.16\\
15330 & 5610 & 107 & 4.552 & 0.058 & 7.24 & 1.23 & 3.48\\
15371 & 5751 & 111 & 4.517 & 0.063 & 7.24 & 1.24 & 3.48\\
15457 & 5564 & 104 & 4.523 & 0.057 & 7.49 & 1.24 & 3.90\\
15510 & 5390 & 106 & 4.578 & 0.051 & 7.16 & 1.17 & 3.30\\
16537 & 5052 & 100 & 4.621 & 0.044 & 7.47 & 1.15 & 2.97\\
16852 & 5914 & 113 & 4.123 & 0.091 & 7.44 & 1.35 & 3.74\\
17378 & 5023 & 419 & 4.145 & 0.181 & 7.62 & 1.04 & 3.07\\
17420 & 4801 & 161 & 4.633 & 0.047 & 7.41 & 1.00 & 2.79\\
19849 & 5164 & 98 & 4.614 & 0.048 & 7.34 & 1.06 & 2.96\\
22263 & 5695 & 106 & 4.504 & 0.058 & 7.45 & 1.25 & 3.46\\
22449 & 6424 & 125 & 4.336 & 0.060 & 7.61 & 1.69 & 9.92\\
23311 & 4641 & 86 & 4.626 & 0.024 & 7.81 & 1.08 & 2.94\\
23693 & 6069 & 117 & 4.451 & 0.064 & 7.35 & 1.46 & 8.90\\
24608 & 5137 & 109 & 2.508 & 0.117 & 0.00 & 0.00 & 0.00\\
24813 & 5781 & 108 & 4.325 & 0.097 & 7.57 & 1.24 & 3.35\\
26779 & 5150 & 95 & 4.584 & 0.041 & 7.66 & 1.08 & 3.24\\
27072 & 6287 & 125 & 4.362 & 0.064 & 7.48 & 1.46 & 5.78\\
27913 & 5820 & 108 & 4.485 & 0.059 & 7.44 & 1.39 & 5.80\\
29271 & 5473 & 102 & 4.507 & 0.059 & 7.58 & 1.01 & 3.43\\
32349 & \dots & \dots  & \dots & \dots  & \dots & \dots & \dots\\
37279 & 6677 & 131 & 4.081 & 0.099 & 7.63 & 1.72 & 4.41\\
37349 & 4889 & 89 & 4.623 & 0.038 & 7.55 & 1.20 & 3.03\\
37826 & 4666 & 95 & 2.685 & 0.091 & 0.00 & 0.00 & 0.00\\
40693 & 5331 & 98 & 4.569 & 0.050 & 7.42 & 1.13 & 2.90\\
41926 & 5210 & 104 & 4.631 & 0.057 & 7.16 & 1.17 & 2.94\\
42438 & 5684 & 116 & 4.514 & 0.061 & 7.36 & 1.43 & 5.86\\
42808 & 4930 & 89 & 4.619 & 0.038 & 7.61 & 1.19 & 3.84\\
43587 & 5063 & 110 & 4.558 & 0.038 & 7.87 & 0.92 & 3.33\\
46853 & 6310 & 122 & 3.913 & 0.089 & 7.42 & 1.71 & 5.70\\
47080 & 5301 & 97 & 4.513 & 0.061 & 0.00 & 0.00 & 0.00\\
          \noalign{\smallskip}
            \hline
         \end{array}
     $$ 
\begin{list}{}{}
\item[$^{\mathrm{a}}$] This iron abundances were derived from the simultaneous
	inversion of all iron lines. These values were adopted for the model atmospheres
	but are not the same as those used in the discussion to follow in
	\S \ref{discussion}, which are based on Fe I lines only.
\end{list}     
   \end{table}

   \begin{table}
      \caption[]{Stars in the sample. (cont.)}
         \label{runs}
     $$ 
         \begin{array}{llllllll}
            \hline
            \noalign{\smallskip}
            {Hipp. \#}  & T_{\mathrm{eff}} &  \sigma  &  {\mathrm{log}g} & \sigma & {\mathrm{[Fe/H]}^{\mathrm{a}}} & \xi  & \eta \\
            \noalign{\smallskip}
	     & ({\mathrm K}) &    & ({\mathrm cgs}) &	&  
	     & ({\mathrm{km~s}}^{-1}) & ({\mathrm{km~s}}^{-1}) \\            
            \hline
            \noalign{\smallskip}
51459 & 6057 & 114 & 4.425 & 0.066 & 7.42 & 1.29 & 3.79\\
53721 & 5751 & 106 & 4.349 & 0.097 & 7.48 & 1.18 & 3.42\\
56452 & 5118 & 97 & 4.637 & 0.054 & 7.14 & 1.14 & 2.76\\
56997 & 5402 & 99 & 4.566 & 0.053 & 7.42 & 1.16 & 3.08\\
57443 & 5558 & 109 & 4.545 & 0.059 & 7.20 & 1.12 & 3.54\\
57632 & \dots  & \dots  & \dots  & \dots  & \dots & \dots & \dots\\
57757 & 6076 & 119 & 4.142 & 0.099 & 7.68 & 1.49 & 3.86\\
58576 & 5361 & 100 & 4.473 & 0.076 & 7.72 & 1.10 & 3.24\\
61317 & 5743 & 109 & 4.470 & 0.071 & 7.26 & 1.22 & 3.19\\
61941 & 6960 & 146 & 4.304 & 0.062 & 7.63 & 1.91 & 15.37\\
64241 & 6347 & 130 & 4.376 & 0.149 & 7.41 & 1.65 & 10.72\\
64394 & 5910 & 111 & 4.436 & 0.064 & 7.54 & 1.29 & 3.86\\
64797 & 4915 & 101 & 4.632 & 0.048 & 7.40 & 1.10 & 3.09\\
64924 & 5483 & 100 & 4.524 & 0.058 & 7.45 & 1.17 & 3.01\\
67422 & 4416 & 141 & 4.619 & 0.020 & 0.00 & 0.00 & 0.00\\
67927 & 5942 & 115 & 3.750 & 0.107 & 7.84 & 1.72 & 7.94\\
68184 & 4647 & 155 & 4.621 & 0.027 & 7.81 & 0.95 & 2.97\\
69673 & 4158 & 127 & 1.887 & 0.164 & 0.00 & 0.00 & 0.00\\
69972 & 4671 & 86 & 4.620 & 0.027 & 0.00 & 0.00 & 0.00\\
71681 & 4970 & 180 & 4.592 & 0.038 & 7.73 & 0.81 & 3.32\\
71683 & 5519 & 123 & 4.256 & 0.103 & 7.67 & 1.04 & 3.71\\
72659 & 5350 & 115 & 4.576 & 0.050 & 7.33 & 1.19 & 3.64\\
72848 & 5115 & 93 & 4.580 & 0.040 & 7.59 & 1.11 & 3.66\\
73695 & 5510 & 108 & 4.109 & 0.100 & 7.12 & 0.94 & 3.18\\
77257 & 5819 & 110 & 4.297 & 0.098 & 7.51 & 1.33 & 3.92\\
77952 & 7384 & 368 & 4.219 & 0.066 & 8.01 & 2.20 & 37.72\\
78072 & 6246 & 122 & 4.296 & 0.087 & 7.40 & 1.52 & 6.55\\
78775 & 5247 & 107 & 4.646 & 0.049 & 6.93 & 1.17 & 2.76\\
79190 & 5049 & 108 & 4.656 & 0.058 & 7.14 & 1.24 & 3.28\\
79672 & 5693 & 108 & 4.478 & 0.062 & 7.53 & 1.16 & 3.67\\
80337 & 5730 & 109 & 4.499 & 0.059 & 7.48 & 1.33 & 3.63\\
80686 & 5987 & 122 & 4.459 & 0.061 & 7.42 & 1.37 & 3.92\\
81300 & 5165 & 93 & 4.587 & 0.041 & 7.56 & 1.17 & 3.11\\
81693 & 5655 & 148 & 3.672 & 0.123 & 7.62 & 1.38 & 4.04\\
84405 & 4960 & 173 & 4.572 & 0.037 & 7.31 & 1.15 & 3.37\\
84720 & 5131 & 106 & 4.614 & 0.046 & 7.15 & 1.05 & 3.32\\
84862 & 5618 & 111 & 4.405 & 0.101 & 7.15 & 1.25 & 3.02\\
85235 & 5240 & 100 & 4.625 & 0.052 & 7.18 & 1.02 & 2.86\\
86032 & \dots  & \dots  & \dots  & \dots  & \dots & \dots & \dots\\
86036 & 5726 & 112 & 4.420 & 0.082 & 7.44 & 1.17 & 3.93\\
86400 & 4833 & 86 & 4.624 & 0.041 & 7.50 & 1.00 & 3.04\\
86974 & 5397 & 107 & 3.965 & 0.094 & 7.80 & 1.04 & 3.44\\
88601 & 5050 & 105 & 4.562 & 0.038 & 7.51 & 0.98 & 3.13\\
88972 & 5000 & 91 & 4.622 & 0.048 & 7.39 & 1.07 & 2.83\\
89937 & 6048 & 128 & 4.336 & 0.102 & 7.04 & 1.02 & 4.26\\
90790 & 4953 & 171 & 4.632 & 0.047 & 7.34 & 1.10 & 2.90\\
91262 & \dots & \dots  & \dots & \dots  & \dots & \dots & \dots\\
91438 & 5524 & 105 & 4.574 & 0.055 & 7.24 & 1.24 & 3.40\\
96100 & 5218 & 96 & 4.611 & 0.049 & 7.33 & 1.07 & 2.84\\
97649 & 7646 & 388 & 4.230 & 0.086 & 0.00 & 0.00 & 0.00\\
97944 & 4600 & 86 & 4.559 & 0.032 & 7.30 & 2.68 & 40.82\\
98036 & 5106 & 183 & 3.541 & 0.119 & 7.36 & 1.15 & 3.48\\
99240 & 5347 & 107 & 4.248 & 0.096 & 0.00 & 0.00 & 0.00\\
99461 & 4953 & 99 & 4.667 & 0.054 & 7.01 & 1.19 & 3.17\\	   
          \noalign{\smallskip}
            \hline
         \end{array}
     $$ 
   \end{table}

   \begin{table}
      \caption[]{Stars in the sample. (cont.)}
         \label{runs}
     $$ 
         \begin{array}{llllllll}
            \hline
            \noalign{\smallskip}
            {Hipp. \#}  & T_{\mathrm{eff}} &  \sigma  &  {\mathrm{log}g} & \sigma & {\mathrm{[Fe/H]}^{\mathrm{a}}} & \xi  & \eta \\
            \noalign{\smallskip}
	     & ({\mathrm K}) &    & ({\mathrm cgs}) &	&  
	     & ({\mathrm{km~s}}^{-1}) & ({\mathrm{km~s}}^{-1}) \\            
            \hline
            \noalign{\smallskip}
99825 & 5022 & 94 & 4.604 & 0.044 & 7.50 & 1.06 & 3.21\\
102422 & 4971 & 412 & 3.540 & 0.180 & 7.41 & 1.11 & 3.19\\
105858 & 6054 & 139 & 4.474 & 0.074 & 6.89 & 1.27 & 3.62\\
107556 & 7317 & \dots  & 4.917 & \dots  & \dots & \dots & \dots\\
109176 & 6455 & 132 & 4.298 & 0.075 & 7.45 & 1.53 & 5.32\\
110109 & 5704 & 112 & 4.488 & 0.072 & 7.21 & 1.24 & 3.44\\
113368 & \dots & \dots  & \dots & \dots  & \dots & \dots & \dots\\
114622 & 4743 & 86 & 4.628 & 0.035 & 7.64 & 1.00 & 3.01\\
116727 & 4625 & 94 & 3.473 & 0.196 & 0.00 & 0.00 & 0.00\\
116771 & 6174 & 118 & 4.181 & 0.104 & 7.46 & 1.39 & 4.71\\
117712 & 4660 & 110 & 4.712 & 0.040 & 6.62 & 0.02 & 4.35\\ 	   
          \noalign{\smallskip}
            \hline
         \end{array}
     $$ 
   \end{table}

\subsection{Kinematics}

Radial velocities were compiled from the catalogs of Barbier-Brossat \& Figon
(2000) and Malaroda, Levato \& Galliani (2001). Proper motions and
parallaxes were extracted from the  {\it Hipparcos} catalog (ESA 1997).
We derived stellar UVW galactic velocities following Johnson \& 
Soderblom (1987), but as all {\it Hipparcos} astrometry is given in the
ICRS system J1991.25 (TT), we use $\alpha_{\rm NGP}= 192.85948$ deg, 
$\delta_{\rm NGP}= 27.12825$ deg, and $\theta_0 = 122.93192$ deg (ESA 1997;
vol 1). We modified Johnson \& Soderblom's formulation to account for
the correlations between the astrometric parameters given in the {\it
Hipparcos} catalog. The details are explained in Appendix B. 
No corrections have been introduced for the solar peculiar
motion. The derived galactic velocities and input data are given in the
electronic Table 5.

\subsection{Chemical abundances}
\label{abu}

We carried out an extensive search for atomic data, mainly transition 
probabilities, for elements producing lines in the spectral range of our
observations. Our search concentrated on reliable 
laboratory measurements and accurate theoretical 
calculations. Candidate lines were examined in the high-dispersion atlases of
 Arcturus (Hinkle et al. 2001), Procyon (Griffin \& Griffin 1979; 
 Allende Prieto et al. 2002c), and the Sun (Kurucz et al. 1984), to identify 
 the best abundance indicators.

A total of 68 features were selected. These features were single-component
lines, or the result of multiple transitions (hfs, isotopic shifts, and blends), 
requiring 275 components and providing abundances for 22 chemical elements. 
Our analysis considers line profiles and,
therefore, we were able to use full profiles or parts of line profiles. 
Natural and van der Waals broadening were accounted for using data from 
VALD\footnote{http://www.astro.univie.ac.at/$\sim$vald/} (Kupka et al. 1999)
and the calculations of Barklem, Piskunov \& O'Mara (2000c, 
and references therein) respectively. 
Hyperfine structure and isotopic splittings were considered for lines of
 Cu I (Kurucz's web site\footnote{http://kurucz.harvard.edu}), 
 Sc II (data compiled by Prochaska \& McWilliam 2000 for $\lambda$6245.622 
 and $\lambda$6604.6; from Mansour et al. 1989 for $\lambda$6300.684 
 and $\lambda$6320.843),  Mn I (compiled by Prochaska \& McWilliam 2000),  
 Ba II (compiled by McWilliam 1998) and Eu II (Lawler et al. 2001). Solar
 isotopic ratios were always assumed.
Table 6, available only electronically, provides a summary of our adopted
wavelengths, excitation potentials, and
transition probabilities.

The  parts of the spectrum selected for the analysis define a 'mask' that we
use for all late-type stars. We exclude the core of strong lines (relative 
flux lower than 60 \% of the pseudo-continuum level), in an attempt to
minimize the impact of departures from LTE. 
We analyze the spectra with MISS  (Allende Prieto et al. 1998, 2001), adopting
theoretical '97 MARCS LTE blanketed 
model atmospheres (Gustafsson et al. 1975; Asplund et al. 1997), to find the
 set of abundances, microturbulence, and macroturbulence 
 that best reproduces the observations. 
%The '97 MARCS models adopt Grevesse \& Sauval  abundances (i.e. a 'low' iron
%abundance)
%Details on the radiative transfer calculations 
%can be found in Allende Prieto et al. (2001). 
We chose to use a Gaussian
broadening function (referred to as `macroturbulence' or $\eta$) to account for 
large-scale turbulence (proper macroturbulence),  
rotational and instrumental broadening. 
The determination of
abundances,  micro, and macroturbulence, 
is accomplished by minimizing the $\chi^2$ between synthetic and observed
line profiles, and this can be carried out with all lines simultaneously, 
or line-by-line. 
After preliminary all-lines runs to determine
the micro and macroturbulence, and the overall metallicity to be adopted in 
the calculation of the model atmospheres, we used a 
line-by-line approach to identify outliers and estimate internal uncertainties.
Table 7, available only electronically, 
provides the final adopted abundances for all stars. These abundances are
discussed in Section \ref{abus}. 
The stated uncertainties are the standard deviation from different lines 
(which is set to zero when there is only one line involved).

The median rms scatter we find for the abundance of a
given element in a given star is 0.04 dex, but
 other systematic contributors exist. Departures from LTE and 
stellar granulation in the real photospheres will bias 
abundance estimates from individual lines differently. We refer every single
abundance estimate from a line to the solar abundance from the same line.
Thus, our approach is \emph{strictly differential}.
This should reduce systematic errors due to the 
shortcomings of our modeling, but given the large
spread in atmospheric parameters of our stars, will certainly not succeed in 
removing them completely. Dealing with NLTE and inhomogeneities 
is very  important in some cases,
%, and will
%likely refine chemical abundances from stellar spectra in the future, 
but exceeds the scope of this work. These errors 
add to the pool  of uncontrolled 
uncertainties together with unrecognized blends, and errors in the 
continuum normalization. Finally, even for these nearby stars, 
%, whose
%properties are known much better than for more distant objects, 
errors in the
atmospheric parameters are a significant, fortunately controlled, source
of uncertainty. We have estimated such errors by repeating the calculations 
with
model atmospheres with perturbed atmospheric parameters. As the gravities and
metallicities are precisely determined and these parameters only produce
minor changes in the abundances compared to variations in the $T_{\rm eff}$, 
we have estimated the effect of increasing the $T_{\rm eff}$s  by 
$3\sigma$ and added that information to Table 7.

\begin{figure}
\centering
{\includegraphics[width=5.5cm,angle=90]{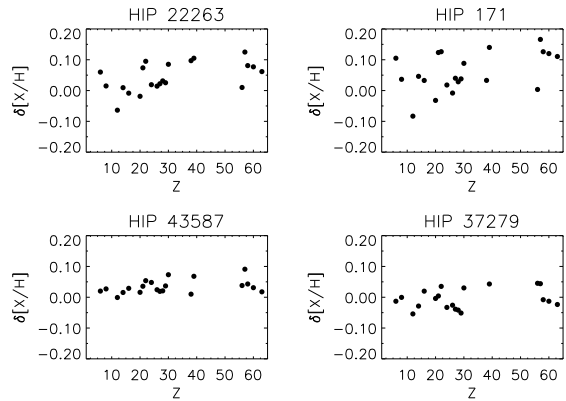}} 
\caption{A sample of the 
change in the abundances as a function of the atomic number 
when Kurucz's (1993) model atmospheres 
replace MARCS models. A solar-like star (HIP 22263),  two late G-type
 stars with extreme metallicities in our sample (HIP 171 with 
 [Fe/H] $\simeq -0.8$
 and HIP 43587 with [Fe/H] $\simeq +0.3$), and an 
 F-type star with solar-like abundances
 (HIP 37279) are shown.}
\label{km}
\end{figure}

Adopting Kurucz (1993) non-overshooting model atmospheres instead
of MARCS models, we systematically
change the temperature and pressure stratification, and then the derived
abundances. Although the changes are generally less than $0.1$ dex, for some
ranges of the atmospheric parameters and some particular elements 
%and for some elements represented by a single ionization stage 
they can be somewhat larger. This is
illustrated in Fig. \ref{km} for a solar-like star (HIP 22263),  
two late G-type stars with extreme metallicities in our sample 
(HIP 171 with [Fe/H]$ \simeq -0.8$
 and HIP 43587 with [Fe/H]$\simeq +0.3$), and an F-type star 
 with solar-like abundances (HIP 37279). 
 
\section{Ionization and excitation balance} 
\label{balance}

\begin{figure*}
\centering
%{\includegraphics[width=5.5cm,angle=90]{./ionization_si.ps}}
%{\includegraphics[width=5.5cm,angle=90]{./ionization_ti.ps}}
%{\includegraphics[width=5.5cm,angle=90]{./ionization_v.ps}}
{\includegraphics[width=8.5cm,angle=90]{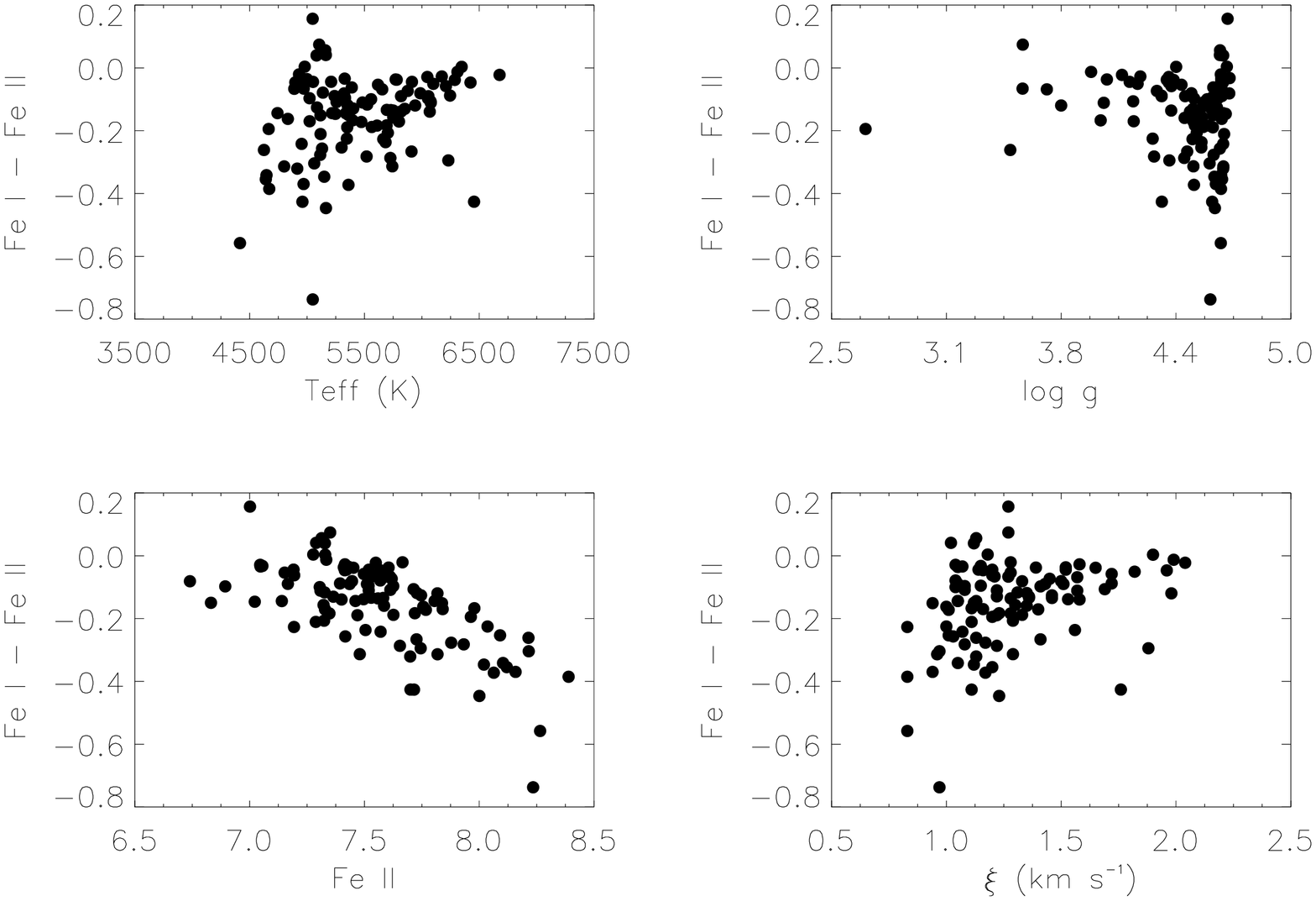}} 
%{\includegraphics[width=5.5cm,angle=90]{./ionization_zr.ps}}
{\includegraphics[width=8.5cm,angle=90]{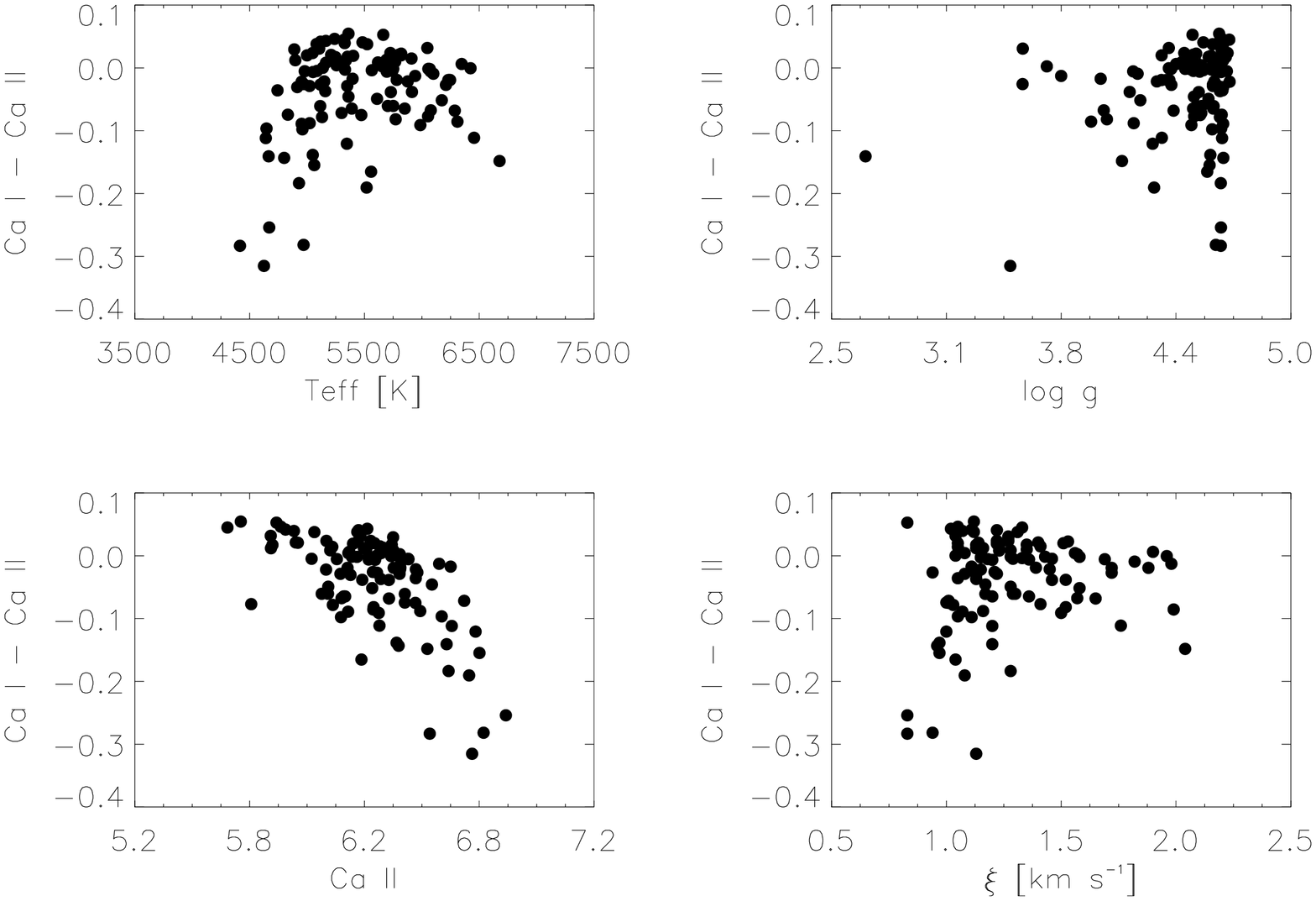}}
\caption{Differences between the mean 
iron abundances derived from lines of different ionization stages of
Fe and Ca for the stars in the sample
as a function of several atmospheric parameters.}
\label{ionization}
\end{figure*}

As a consistency check of the high-dispersion spectroscopic analysis,
we examine whether the abundances derived from different lines of the 
same element agree. 
Inspection of the abundances  from lines of different 
excitation for
Fe I (Ti I, or Ca I) reveals no clear sign of correlations. 
However, it soon 
became obvious that iron abundances derived from lines of neutral or ionized
iron provided systematically different values. This effect is illustrated
in Fig. \ref{ionization} for Fe and Ca, where it can be seen that the 
differences can reach $\sim 0.5$ dex for the most metal-rich stars 
in the sample. Similar effects are apparent for Si and Ti. 
A parallel analysis of the nearby star sample 
based on Kurucz's model atmospheres shows a similar 
pattern and magnitude for the ionization imbalance to that
 found with MARCS models,  
although the detailed shape of the trends in Fig. \ref{ionization} changes.

\subsection{Precedents}

Serious worries about the ionization balance in K-type
dwarfs appeared in the work by 
%by Cayrel de Strobel (1966), Strohbach (1970) and
%particularly 
Oinas (1974). Perrin, Cayrel de Strobel \& Cayrel (1975) 
revisited the issue and concluded that the magnitude of the problem
 was overestimated by Oinas
and the discrepancies perhaps not significant.

In more recent times, Feltzing \& Gustafsson (1998)
 reported the case of five K-type dwarfs
 %, two of them 
% (HIP 37349 $\equiv$ HD 61606 A and 
% HIP 23311 $\equiv$ HR 1614 $\equiv$ HD 32147) 
% in common with our sample, 
that showed dramatic differences between the 
 abundances from lines of neutral and ionized Fe, and the same effect was
 apparent for Ca and Cr. 
 Thor\'en \& Feltzing (2000) reanalyzed some of the problematic stars in
 Feltzing \& Gustafsson (1998), suggesting adjustments to the
 the $T_{\rm eff}$ scale to solve the discrepancy. 
 
 Very recently, Schuler et al. (2003) have reported similar problems with
 several K dwarfs in the open cluster M34. With effective temperatures
 set to satisfy the excitation equilibrium of Fe I, and gravities 
 assigned from the positions of the stars on the main sequence, 
 their standard
 LTE analysis resulted in consistent abundances from Fe I and Fe II lines
 for warm stars (T$_{\rm eff} \ge 5500$ K), but significant deviations
 from the ionization balance for most of the cooler dwarfs. Similar to
 the cases of 
 the nearby K dwarfs and those in the Feltzing \& Gustafsson samples, 
 Schuler et al. found significantly larger abundances (up to 0.6 dex) 
 from Fe II lines. The abundances they derived 
 for the K-type stars  from Fe I lines 
 were similar to the abundances for warmer cluster stars that
  did not show the problem.
  The symptoms are not unique to M34, as shown by Yong et al. (2004) 
for the Hyades. The iron abundance inferred from Fe II lines
continues to diverge at cooler effective temperatures, exceeding
the values from Fe I lines by more than an order of magnitude for 
stars with $T_{\rm eff} \sim 4000$ K in this cluster.

As we discuss below, adjusting the stellar 
 parameters to make ends meet is not easily justified for our sample.
 We should note that most of the previous high-resolution
 spectroscopic surveys have been for warm (F- and 
 early G) stars with  modest iron enrichments  
 (see, e.g. Edvardsson et al. 1993;  Reddy et al. 2003). From inspection of 
 Fig. \ref{ionization} we would expect those studies to find only minor
 ($\la 0.2$ dex) deviations from the LTE ionization balance.

\subsection{Errors in the atmospheric parameters?}
 
 To reconcile the abundances from neutral and ionized lines for the most 
 metal-rich (and cool)  stars in Fig. \ref{ionization} by modifying the
 trigonometric gravities would require to decrease them from 
 $\log g \simeq 4.6$
 to $\sim 3.8$. Similarly, the abundances could be reconciled by increasing
 the $T_{\rm eff}$s by $\sim 350$ K.
Our list of Fe I was punctiliously selected for being clean and having
accurate atomic data, but it spans a limited range in excitation 
energy. 
A test for HIP 69972 ($T_{\rm eff} \sim 4700$ K) with 47 Fe I lines
drawn from the lists of Blackwell, Lynas-Gray \& Smith (1995) and 
Holweger,  Kock \& Bard (1995)
%\footnote{See also references by the same groups on iron}
spanning a range of excitation potential from 0.1 to 4.7 eV 
shows that a change of 300 K (from 4671 K to 4971 K) results in 
a change in the slope of the abundances from 0.06 to $-$0.01 dex/eV, 
increasing the mean abundance from Fe I by only 0.05 dex, 
and bringing a marginal improvement in the standard deviation 
from 0.14 dex to 0.13 dex. 
The same change is able to bring down the 
mean abundance derived from Fe II lines by roughly 0.5 dex. 

\begin{figure}
\centering
{\includegraphics[width=8.5cm,angle=0]{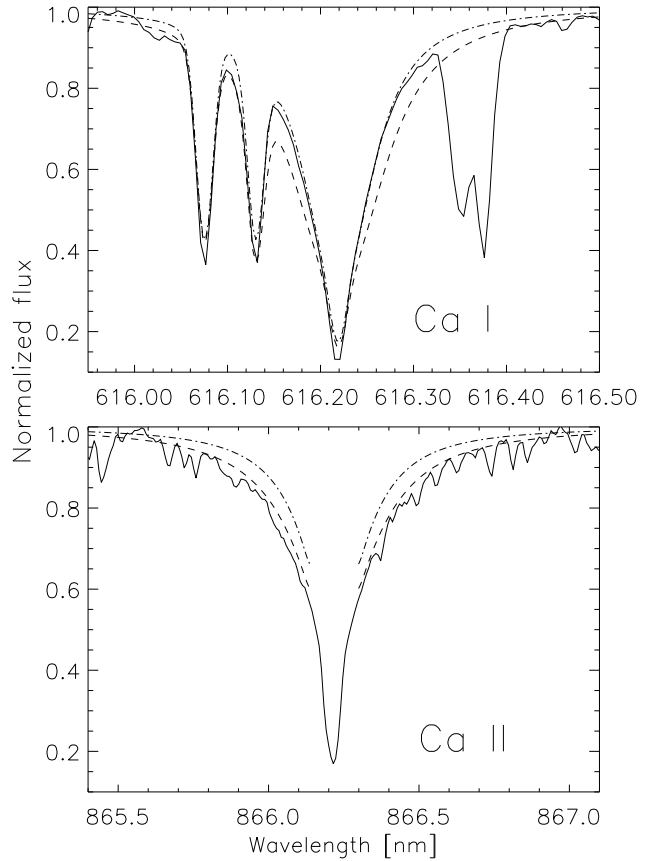}}
\caption{The Ca I line at 616.22 and the Ca II line at 866.22 observed
in the spectrum of HIP 69972 (solid lines) point to different abundances.
The dashed and dash-dotted lines are spectral synthesis for the 
same appropriate stellar parameters but abundances that differ by 0.25 dex.}
\label{strong}
\end{figure}

The lines of neutral metals with weak-to-moderate strength in a solar like
star tend to become very strong in cool metal-rich stars. With stronger lines
the role of the damping constants is enhanced (see, e.g., Ryan 1998). Our
analysis makes use of line profiles, rather than equivalent widths, and
adopts, when available, 
the damping parameters calculated by Barklem et al. (see \S 3), which
we consider reliable to within $\sim$ 10\%. Our neutral iron lines 
are significantly stronger (equivalent widths between 7 and 14 pm) 
for a star like HIP 69972 than in the Sun. The selected Fe II lines, in
turn, are all weaker than 8 pm in the spectrum of HIP 69972. 
The situation for our calcium lines, however, is quite different, and yet 
the same shift between the abundances from lines of neutral and ionized 
species is found. As shown in Fig. \ref{strong}, even if we only consider
the wings of strong Ca lines in the analysis, Ca I and Ca II lines point
to quite different abundances. These differences could again be bridged by
reducing the surface gravity by $\sim 0.6$ dex, or increasing the effective 
temperature by $\sim 280$ K (weakening the Ca I line).

\begin{figure}
\centering
{\includegraphics[width=8.5cm,angle=0]{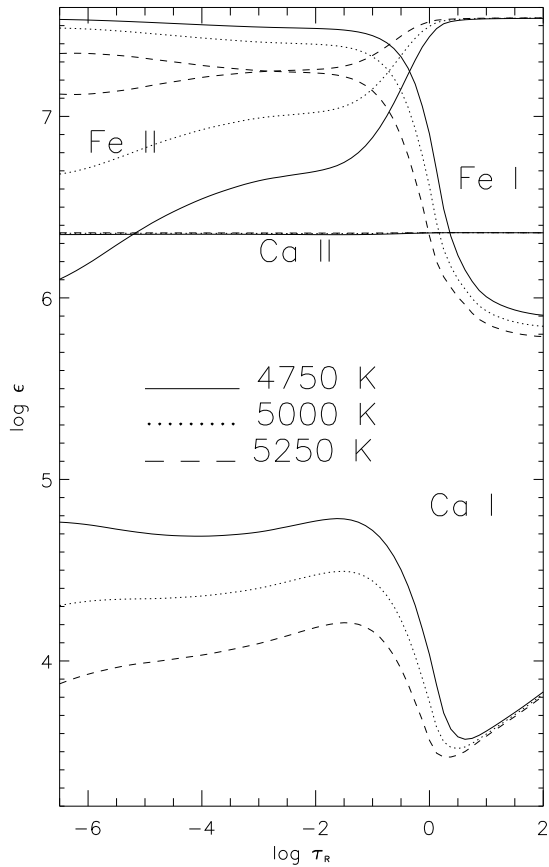}}
\caption{Logarithm of the ratio between the abundance of 
neutral or ionized metals and hydrogen atoms for three model atmospheres 
with $T_{\rm eff} = 4750$ K (solid line), 5000 K (dotted line), and 5250 K
(dashed line). The optical depth scale is based on Rosseland's mean
opacity.}
\label{transition}
\end{figure}

However, a systematic modification by 
 $\sim 300$ K of the effective temperature from the IRFM calibrations
well exceeds the expected uncertainties. 
For example,  the $T_{\rm eff}$ calibration 
based on $(B-V)$ for dwarfs and subgiants, was derived from data for
 410 stars, yielding a standard deviation of 130 K. The calibration for the
 Str\"omgrem photometry was based on 289 stars with a standard deviation
 of 141 K.
 In addition, we find an excellent agreement between the
$T_{\rm eff}$s derived from $(B-V)$ and Str\"omgrem photometry, 
and, independently, from
the analysis of the wings of Balmer lines (see \S \ref{pars}). 
Finally, we should emphasize that the stellar parameters for our sample
are well bracketed by those of the stars used to define 
the Alonso et al. calibrations.

It is intriguing to consider that the problem may be associated with the
trigonometric gravities, 
but this is extremely unlikely given the size of the discrepancy.
From inspection of Tables 2-4, one notices that the corrections required are
about 10 times the estimated uncertainties in this parameter.
Additionally, Feltzing \& Gustafsson
arrived at similar gravities for their K dwarfs
(in some cases very different from those
determined spectroscopically) from trigonometric parallaxes and the 
pressure-sensitive wings of strong metal lines. Interestingly, 
 comparison between precise stellar radii for stars in eclipsing binaries 
 with evolutionary isochrones has shown that G and K-type stars in 
 detached systems are brighter than predicted (see, e.g., 
 Clausen et al. 1999; Allende Prieto 2001; Torres \& Ribas 2002;
 Ribas 2003). For a reliably determined $M_V$, 
 predicted radii are smaller by $\sim$ 10 \% and therefore gravities
 larger by $\sim$ 0.1 dex. This effect is as yet not well 
 understood (see, e.g., 
 Lastennet et al. 2003 for a possible solution), 
 but cannot explain the ionization shift we find, because it is too small.

As tempting as it may be to modify the stellar parameters to comply with
the ionization balance, we cannot do so without violating robust
 observational constraints. 
  
\subsection{Possible explanations}

Ionization shifts with respect to theoretical predictions
have been previously reported for metal-poor stars
in the form of differences between spectroscopic and trigonometric gravities, 
% (e.g. Allende Prieto et al. 1999), 
 and are often attributed to departures from LTE. 
In fact, deriving higher abundances from Fe II lines
seems consistent with the hypothesis of non-LTE overionization
%  (the main non-LTE effect for iron in a solar-like star) 
 as a plausible explanation. 
 The mild correlation of the 
effect with $T_{\rm eff}$ in Fig. \ref{ionization} suggests that the
most serious discrepancies 
appear at $T_{\rm eff} \la 5000$ K, which is where
 most of the iron becomes neutral in the line formation region, 
 as shown in Fig. \ref{transition}.
Departures from LTE are believed to cause an overionization of iron 
compared to LTE predictions 
in solar-like stars (see, e.g. Shchukina \& Trujillo Bueno 2001). 
The effects are felt in the analysis of Fe I lines, but not for 
Fe II lines for a solar-temperature dwarf, because most of the iron atoms
are ionized in the line-forming region. However, the argument 
reverses for stars with an effective temperature lower than about 5000 K.
Besides, 
a higher metal abundance in the photosphere decreases gas pressure and 
enhances electron pressure, shifting the Saha ionization balance towards
neutral iron. Thus, departures from LTE in Fe II levels 
would qualitatively agree with the inconsistencies
shown in Fig. \ref{ionization}. However, as shown also in Fig. \ref{transition},
the situation for Ca I, which is ionized with only 6.1 eV in comparison
to 8.9 eV for Fe I, is very different. Ca is still completely ionized 
everywhere in a solar-composition model atmosphere even for an $T_{\rm eff}$ 
as low as 3700 K. Yet, calcium shows a similar ionization problem as iron.

Other factors, such as surface inhomogeneities (granulation, spots, etc.), 
incomplete opacities (due to,
e.g., molecules), chromospheric or magnetic activity,
 may be responsible instead.
In an early K dwarf, the region where the largest velocity and
temperature amplitudes of granulation take place is slightly beneath the
visible layer (Nordlund \& Dravins 1990). 
As discussed by Dravins \& Nordlund (1990a, 1990b) 
based on 3D hydrodynamical 
simulations, this results in relatively small 
line asymmetries and shifts compared to warmer stars, but causes 
the physical conditions to change very rapidly with the formation 
height of a line.
The formation of high excitation 
lines may be biased towards the hotter granules, making the lines
of ionized species stronger than predicted with a homogeneous model.

In any event, it appears safer in the context of this study  
 to embrace the abundances from the dominant 
species for the cooler stars (which are also the more metal-rich stars) 
of our sample. As the
abundances from Fe I and II lines are roughly in agreement for most of the
warmer stars, in what follows we adopt the mean iron abundance
derived from Fe I lines as our metallicity indicator. We still will use
line-to-line differential abundances with respect to the Sun, and therefore
the notation [Fe/H] will, from now on, refer to Fe I  abundances
relative to the Sun.

\section{Discussion}
\label{discussion}

\subsection{Selection of 'normal' stars}

Some of the observed stars rotate too fast to be handled similarly to the
rest of the sample and were not included in the subsequent analysis. 
These are: HIP 32349, HIP 57632,  HIP 77952, 
HIP 86032,  HIP 91262, HIP 97649, HIP 107556, and  HIP 113368.  HIP 24608 
was found to have peculiar profiles and should be studied separately,  and 
HIP 61941  is a visual binary with a separation of about 1.6 arcsec
 that could not be resolved at the time of the observations.
 HIP 69673, Arcturus, is the star with the lowest gravity in the sample. 
 Its surface composition does not match that of typical disk stars
  (Peterson et al. 1993), and was also excluded.
 
Spectroscopic binaries were not
set aside a priori, unless double lines were obvious. 
The doubled-lined
K-type spectroscopic binaries HIP 97944 (HR 7578) and HIP 117712 (HR 9038) 
will not
be discussed here. Both systems were observed when the lines are
well separated.  
HR 7578 has a period
of 46.8 days, and an eccentricity of 0.68 (Fekel \& Beavers 1983). It is
known to exhibit peculiar abundances (Taylor 1970; Oinas 1974). HR 9038 is
part of a multiple system, with at least a nearby M2. It has a period of
7.8 days and a circular orbit (Halbwachs et al. 2003), but no abundances
have been reported before.  The spectra of these systems deserve a more 
detailed study which would depart from the scope of this paper. 

Fast rotators and single-lined binaries 
are included in the discussion of kinematics. 
Ignoring the possibility of cleaning our sample with
external information will allow us to perform a meaningful comparison
with larger samples of more distant, and therefore less-studied,
stars (see Section \ref{extended}).
Even if the multiplicity fraction is as high as 
$10-20$ \% (Halbwachs et al. 2003; Mason et al. 2003), this will
have only a limited effect on the derived velocity distributions. 
Fast rotating and double-lined binaries discussed above are excluded
from the abundance analysis, but 
single-lined spectroscopic 
binaries with apparently clean spectra are not, 
as no significant differences were noted between those 
identified as members of binary systems in SIMBAD and the rest (see below).

\subsection{Kinematics and metallicity distribution}

\begin{figure}
\centering
{\includegraphics[width=8.5cm,angle=0]{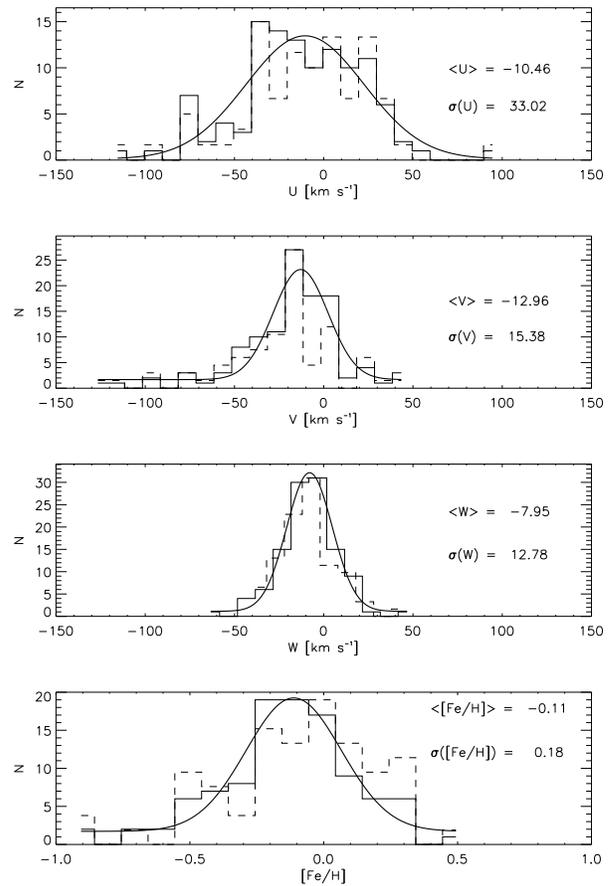}} 
\caption{Histograms of the galactic velocity components (UVW) 
and the metallicity of the sample stars.
The distributions for the complete sample are shown with solid lines. Gaussian
curves have been fitted to these distributions and are also shown with 
continuous solid lines. The dashed-lines show the distributions for the
stars with $T_{\rm eff}$ cooler than 5500 K, which have been re-scaled to 
match the maxima of the all-stars distributions.}
\label{kin}
\end{figure}

In Fig. \ref{kin} we show histograms of the galactic velocity components
and iron abundances for the sample. Gaussian distributions have been fit
to the data, and are usually  fair approximations. 
The zero points of the velocity components  reflect the solar
 motion with respect to its nearest neighbors (what has been
sometimes termed the solar {\it basic} motion). These numbers are
in good agreement with older determinations (see, e.g. Delhaye 1965; 
Woolley 1965)
and for U and W, but not for V
due to the dependence of this parameter on the stellar population 
(asymmetric drift), they also agree very well with the
solar peculiar motion with respect to the local standard of rest (circular
motion; see Dehnen \& Binney 1998).
The widths of the distributions are consistent with values derived from
larger samples, but we will revisit this issue in 
\S \ref{extended}.
To explore the effect of biases due to evolutionary effects
(more massive stars that have already died), we have
restricted the sample to stars with $T_{\rm eff}$ cooler than 5500 K. Such
distributions (dashed lines) have been rescaled to match the maxima of
the total distributions and help the eye in the comparison.  No 
significant differences
are apparent in the velocities.

The metallicity distribution of nearby late-type stars has been 
extensively discussed in the literature. This is the first time, however,
that it has been derived for an absolute magnitude-limited sample using
spectroscopy. A Gaussian provides a fair approximation to the [Fe/H]
distribution of our sample. While the width of the distribution is in good
agreement with most previous results, the center of the Gaussian,
 at  $-0.11$ dex, is slightly higher than most studies (but see
 Haywood 2001, 2002). Restricting the analysis to
 the coolest stars ($T_{\rm eff} < 5500$ K), the resulting metallicity 
 distribution is shifted  to higher values. 
 Further reduction of the maximum 
  $T_{\rm eff}$ to $\sim 5000$ K does not alter the [Fe/H] distribution 
  appreciably. 
 The discussion on the ionization balance
  in \S \ref{balance} may leave some readers worried about how uncertainties
  in theoretical model atmospheres might affect the derived distribution.
It suffices to say that adopting the average (Fe I + Fe II) abundances, would 
shift the center of the (full sample -- solid line)
[Fe/H] distribution in the Figure to $-0.06$ dex.
%, and using only Fe II
%abundances, to $+0.06$ dex.
%The iron abundances derived from the automated technique described
%by Allende Prieto (2003) based on fitting the region around H$\beta$, provides
%a distribution centered at [Fe/H] $= 0.02$ with a $\sigma \simeq 0.17$ dex.

\begin{figure}
\centering
{\includegraphics[width=8.5cm,angle=0]{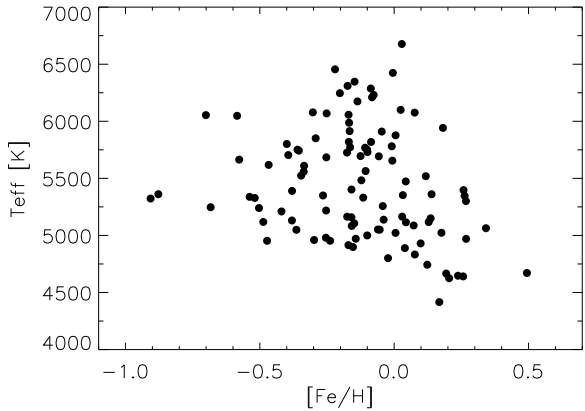}} 
\caption{Fe I  metallicities and effective temperatures for the stars
in our sample.}
\label{fehteff}
\end{figure}

The existence of super-solar metallicity stars in the local disk 
has been known for a long time (e.g. Spinrad 1966; Spinrad \& Taylor 1969; 
Taylor 1970). Curiously, the most metal-rich stars are mainly
dwarfs and subgiants 
with spectral types F or later (e.g. Feltzing, Holmberg \& Hurley 2001;
Allende Prieto 2002b; Taylor 2002).
Fig. \ref{fehteff} shows clearly that the most
metal-rich stars in our volume-limited sample are also the coolest.
Metal-poor main-sequence stars are less luminous than their metal-rich
counterparts, and 
in Figure \ref{fehteff}, the lack of metal-poor stars at the lowest 
temperatures is, at least in part, the result of having a $M_V$-limited 
sample.

HIP 69972 (HD 125072) appears to be the most metal-rich star in the sample. 
The Fe I-based metallicity is roughly $+0.5$ dex,  among the 
largest values reported in the literature (e.g. Buzzoni et al 2001; Taylor 2002;
Chen et al. 2003).
The Fe II-based metallicity of this star is about one dex higher than solar
but,  given the inconsistency between Fe I and Fe II abundances, and the much
higher sensitivity of the Fe II lines to temperature changes, we think such
a higher abundance is unlikely.
This star was previously analyzed by Perrin, Cayrel de Strobel 
\& Dennefeld (1988), who derived a lower value of [Fe/H] $=+0.26$ 
dex\footnote{Their differential analysis used sunlight spectra obtained with
the same instrument as the stellar spectra. Therefore, the quoted abundance
is independent of the adopted iron solar abundance.}.
Reinforcing the discrepancy, 
their adopted effective temperature was nearly 300 K warmer than our 
IRFM value. At least part of the difference can be attributed to their
Fe~I equivalent widths, which we found, on average, to be $\sim $ 30 \% 
smaller than those we measure in the modern CCD spectrum. 
%The equivalent
%width of the only Fe II line they analyzed for this star is, however, in 
%reasonable agreement with our measurement in the ESO spectrum.
% (but they don't use FeI/FeII for logg)
Such a large discrepancy is unexpected, given that the reticon 
spectrum of Perrin et al. had a S/N of 60, and a resolving power 
similar to ours. Variability could be a possible explanation, but 
the {\it Hipparcos} epoch photometry of the star shows a negligible 
broad-band variability of just 0.015 mag over 2.5 years.

%Fig. \ref{slices}  shows the velocity field projected on  
%a Cartesian system  centered at the solar barycenter  
%system at J1991.25 (TT) and with the axes pointing toward the Galactic center,
%the direction of  Galactic rotation and the north Galactic pole. 
%This is the same  that we used for the velocities and therefore 
%$\dot{\rm X} = {\rm U}, \dot{\rm Y} = {\rm V}, \dot{\rm Z} = {\rm W}$. 
%We have subtracted the mean
%velocities shown in Fig. \ref{kin}.
%The lengths of the arrows are arbitrarily normalized and scaled proportionally
%to the modulus of the projected velocity.

%\begin{figure}
%\centering
%{\includegraphics[width=8.5cm,angle=0]{./xy.ps}} 
%{\includegraphics[width=8.5cm,angle=0]{./xz.ps}} 
%{\includegraphics[width=8.5cm,angle=0]{./yz.ps}} 
%\caption{Projections of the velocities in a Cartesian reference frame
%aligned with the Galactic system at the solar position.}
%\label{slices}
%\end{figure}

\begin{figure*}
\centering
{\includegraphics[width=10.2cm,angle=90]{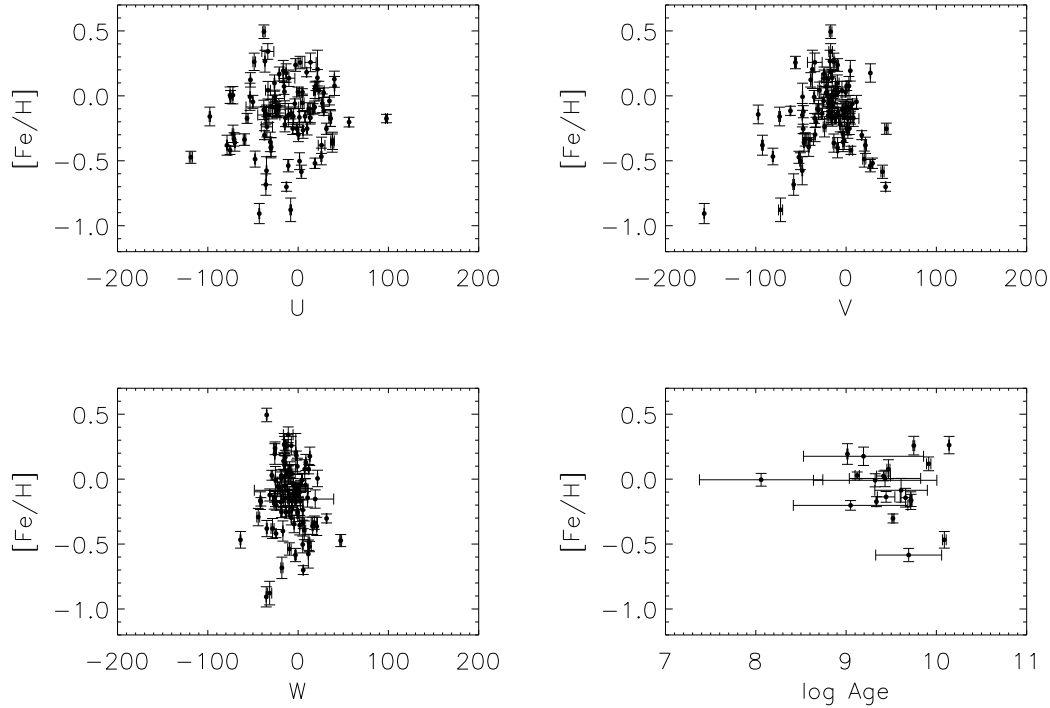}} 
\caption{Distribution of the sample stars in the planes of galactic velocities
(UVW) and [Fe/H], and in the plane of Age vs. [Fe/H].
In the lower-right panel, only stars with relatively
small errors in age ($\sigma \leq$  0.7 dex) are shown. The data point with
a particularly slow galactic rotational velocity $V$ corresponds 
to HIP 5336 (HR 321; $\mu$ Cas),
a well-known metal-poor spectroscopic binary  
(see, e.g., Haywood, Hegyi, \& Gudehus 1992). One of the radial 
velocities in the Malaroda et al. catalog for this star 
is very different from the other two available measurements, 
which is surprising given the small semi-amplitude of the primary (the
secondary is $\sim$ 6.5 mag fainter).
We adopted a recent and very precise  determination of
the systemic velocity
($V_R= -97.44 \pm 0.07$ km s$^{-1}$; T. Forveille, private communication)
which is in good agreement with the other  two measurements
 in the catalogs.}
\label{correla1}
\end{figure*}

Interesting correlations between abundances and kinematics are apparent  
in Fig. \ref{correla1}.
The velocity components of the stars show the smallest scatter at
high iron abundances. The figure for the $V$ component shows an unusually
low density of stars with low metallicity and solar-like galactic rotation:
the most metal-poor stars either lag or lead the typical thin-disk
rotation velocity, with no half measures, but this may be a selection effect
introduced by poor statistics. The lower-right panel in the Figure shows
the age metallicity relation for the stars whose age could be estimated with
an uncertainty of 0.7 dex or less.  Ages could only be determined for the 
few stars visibly
evolved from the main sequence, but this is still quite informative.
This panel warns against interpreting
[Fe/H] as a clock.

Although the most metal-poor stars are also among the 
oldest ($> 10$ Gyr), they share ages with some of the most metal-rich stars
in the sample, which show thin-disk kinematics. 
%This is annoying, as thinking of those metal-rich 
%stars as part of the oldest population of the thin disk, we would have 
%naively expected them
%to be more metal deficient than younger stars with similar kinematics. 
Metal-rich subgiants with old ages have been known for a long
time (see, e.g. Schwarzschild 1958; Eggen 1960) and have been 
 confirmed in post-Hipparcos studies of the  
local disk (e.g. Feltzing \& Holmberg 2000; Feltzing et al. 2001).
Recently, Sandage, Lubin \& Vandenberg (2003) have found that 
evolutionary tracks with a metallicity of  $\sim +0.37$ and an age of
$\sim 8$ Gyr were needed to
reproduce the reddest subgiants in a Hipparcos color-magnitude diagram.
Chen et al. (2003) report on spectroscopy of 
several metal-rich stars with ages $> 10$ Gyr.
We estimate, for example, 
an age of $\sim 13$ Gyr and a metallicity of $+0.3$ for  HIP 99240.
Some of the coolest stars in our sample appear to be slightly off
 the main sequence, and they are most likely binaries. 
 The most notable case is HIP 97944, a BY Dra variable clearly noticeable
 in Fig. \ref{parameters}.
 %, which exhibits strong Ca H and K 
%emission as expected for a young star.

\subsection{Metal abundance ratios}
\label{abus}

\begin{figure*}[t!]
\centering
{\includegraphics[width=12.cm,angle=0]{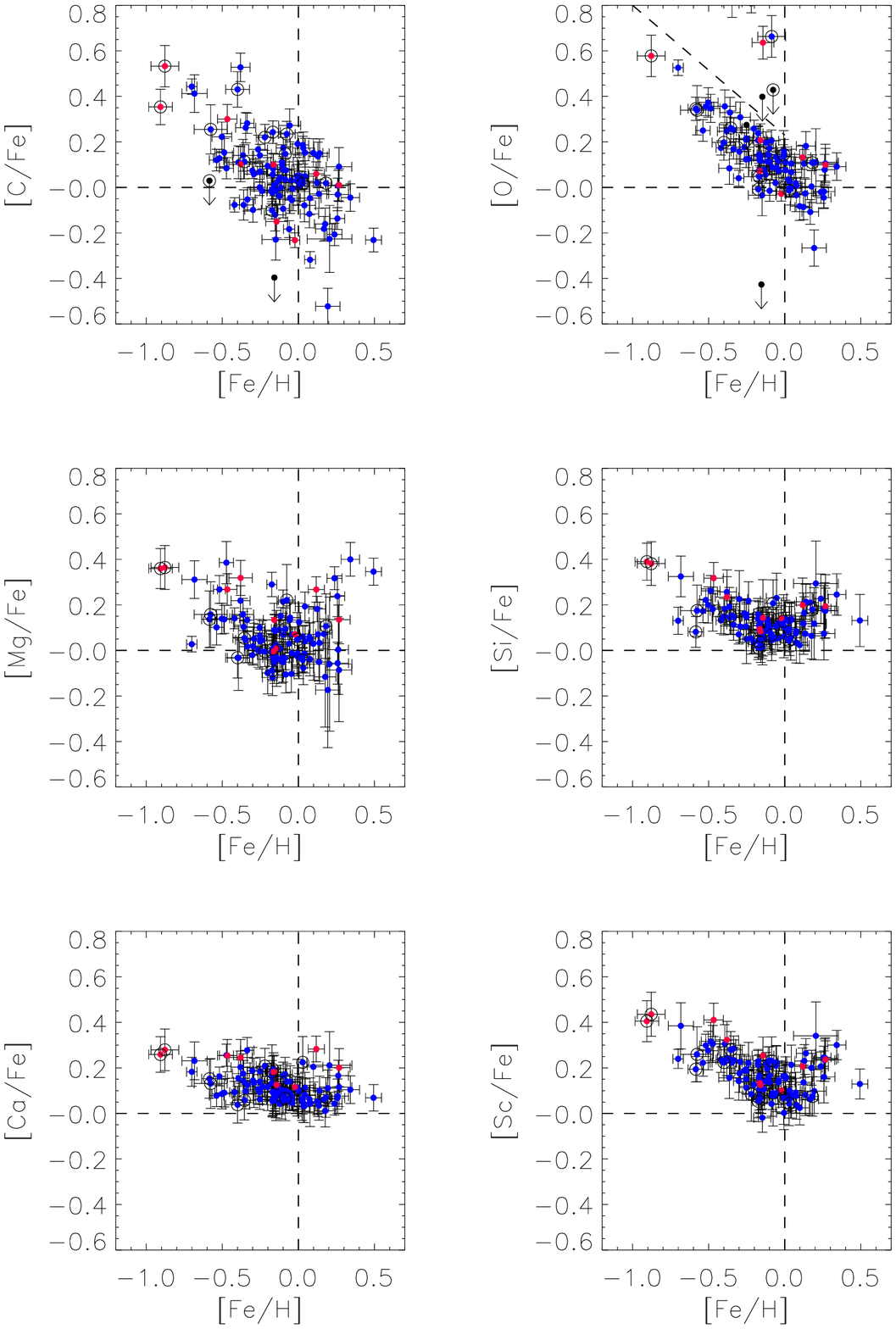}}
%{\includegraphics[width=12.cm,angle=0]{./xfe_c_b.ps}}
%{\includegraphics[width=12.cm,angle=0]{./xfe_c_c.ps}}
%{\includegraphics[width=12.cm,angle=0]{./xfe_c_d.ps}}
\caption{Abundance ratios between different elements and iron as a function
of the iron abundance. Stars with a galactic rotational velocity  
V$<-62$ km s$^{-1}$ (or $<-50$ km s$^{-1}$ 
relative to the local standard of rest)
are shown in red; all others in blue.
Stars labeled in SIMBAD as spectroscopic binaries are
 identified with open circles.}
\label{xfe}
\end{figure*}

\begin{figure*}
\centering
%{\includegraphics[width=12.cm,angle=0]{./xfe_c_a.ps}}
{\includegraphics[width=12.cm,angle=0]{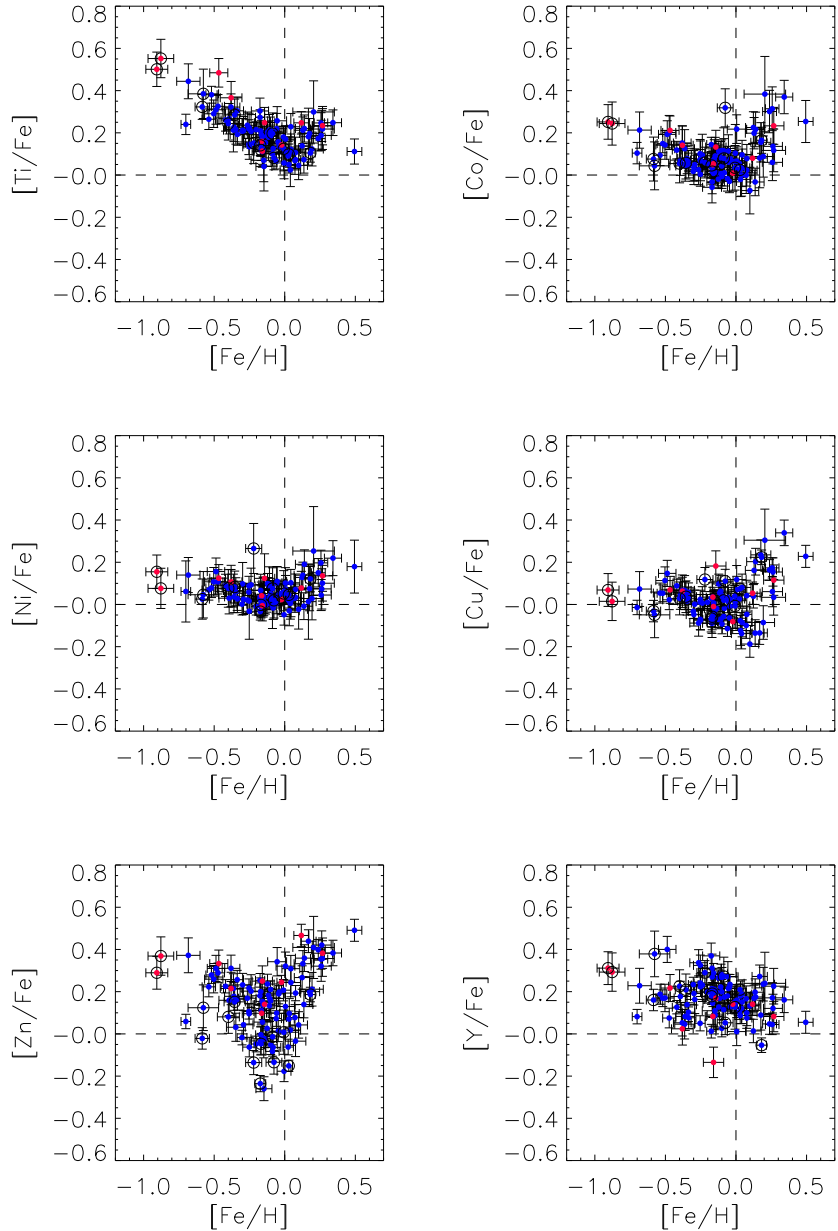}}
%{\includegraphics[width=12.cm,angle=0]{./xfe_c_c.ps}}
%{\includegraphics[width=12.cm,angle=0]{./xfe_c_d.ps}}
\caption{As in Fig. \ref{xfe} for other elements.}
\label{xfe2}
\end{figure*}

\begin{figure*}
\centering
%{\includegraphics[width=6.cm,angle=0]{./xfe_c_a.ps}}
%{\includegraphics[width=6.cm,angle=0]{./xfe_c_b.ps}}
%{\includegraphics[width=12.cm,angle=0]{./xfe_c_c.ps}}
{\includegraphics[width=12.cm,angle=0]{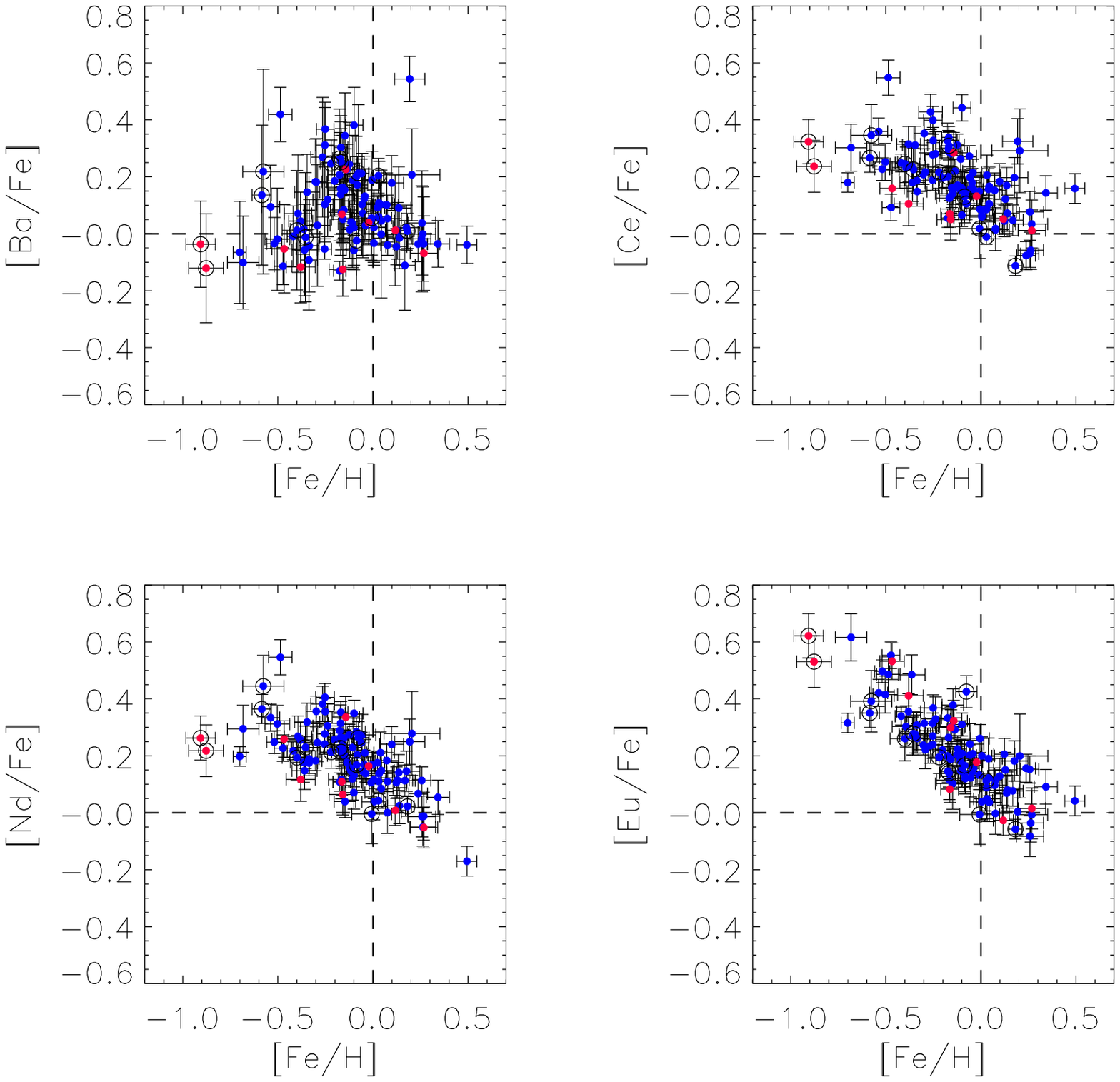}}
\caption{As in Fig. \ref{xfe} and \ref{xfe2} for other elements}
\label{xfe3}
\end{figure*}

The abundance ratios between iron and other metals have been shown to
correlate with the iron abundance for many elements. Such correlations 
provide a rich database to confront with models  of chemical evolution and
nucleosynthesis predictions. 
Because the derived 
trends  may be highly
sensitive to modeling errors (see \S \ref{balance}), 
we have chosen to  restrict our
study to elements represented in the spectrum with low
excitation lines from majority species. Such lines are only weakly
sensitive to changes in the adopted model atmosphere and stellar parameters.
With this filtering, we limit significantly the number of chemical 
elements to explore, but we feel that given the apparent inconsistencies
discussed earlier, this is the only feasible strategy
 to obtain meaningful results.
Figures \ref{xfe}, \ref{xfe2}, and \ref{xfe3} show the ratios of 16 
elements to iron in our sample.
Stars with a galactic rotational velocity  
V$<-62$ km s$^{-1}$ (or $<-50$ km s$^{-1}$ 
relative to the local standard of rest)
are shown in red; all others in blue.
Stars labeled in SIMBAD as spectroscopic binaries are also identified with 
open circles. 
%A null slope in these graphs indicates that the enrichment in a given
%elements follows that in iron. A slope of $\simeq -1$, which may be the 
%case for K and Eu,  indicates that the 
%abundance of the element is not changing as iron increases.
%Many elements in Fig. \ref{xfe} show a monotonic 
%decrease in their abundance ratio to
%iron as the iron abundance increases. 
%Others show a flat trend for [Fe/H]$< 0$, and a decrease at super-solar
%metallicities.
%Distinct trends are apparent for some particular elements. 
%If, as suggested by Fig. \ref{correla1},
%some of the most metal-rich stars are as old as the most metal-poor stars
%in the sample, it should not be surprising that some ratios show a symmetry
%with [Fe/H], as appears to be the case for the $\alpha$ elements or Zn.

{\bf Carbon} is represented in our analysis by 
the forbidden line at 8727 \AA.  This feature
is affected by neighbor lines, 
but we believe they can be properly accounted
for at solar $T_{\rm eff}$s (Gustafsson et al. 1999; 
Allende Prieto et al. 2002d). A straight least-squares linear fit to the data
in Fig. \ref{xfe} (top left panel) suggests 
a slope of $-0.4 \pm 0.1$ dex/dex, significantly steeper that the results of
Gustafsson et al. (1999) for a sample
 of disk stars  spanning a similar range in metallicity. Our C/Fe ratios
 exhibit more scatter, most likely due to the larger range in
 stellar parameters of our nearby stars. Given the
weakness and location of the  forbidden line  
in a crowded region, further analysis will benefit from a
higher resolving power to clarify this issue.

 {\bf Oxygen} is represented very
sparsely in stellar spectra. The only features we use to derive
abundances and discuss here are 
the [O I] line at 630 nm, and the high excitation IR triplet at 770 nm.
While the forbidden line has been shown to be nearly immune to departures from
LTE, it is weak and blended with a Ni I line (Allende Prieto et al. 2001;
Asplund et al. 2004).
The near-infrared triplet lines are an extreme example of departures 
from LTE, but are stronger and clean for 
measuring in solar-like and metal-poor stars.
The high excitation of the triplet lines also argues against their use  
for our study, given the problems with the ionization balance 
(see \S \ref{balance}).
In fact, despite our rigorously differential analysis, we notice that
an analysis of the triplet lines for 
stars with atmospheric parameters close to solar, suggests that
the Sun is oxygen poor by $\sim 0.2$ dex. 

Some insight can be obtained from 
a comparison between the triplet-based and [O I]-based [O/Fe] 
for a solar-like star such as HIP 7918. We have derived an $T_{\rm eff}$  of
$5768 \pm 107$ K, a $\log g$ of $4.422 \pm 0.079$, and a metallicity of
[Fe/H] $= -0.11 \pm 0.07$. The micro and macroturbulence, as well as the
projected rotational velocity are also close
to solar. Yet, analysis of the [O I]- Ni I blend leads to an abundance
[O/H]=0.04, while analysis of the infrared triplet leads to [O/H]$=0.20$.
If departures from LTE in the triplet lines are responsible for this 
discrepancy, they must depend on parameters other than those  
considered in a standard analysis (age is a candidate; see \S \ref{extended}). 
An alternative explanation could be related
to the fact that we cannot derive the solar parameters in the same fashion
as those for the other stars. The trend of O/Fe  in Fig. \ref{xfe} has
been defined from the analysis of the forbidden line. 
We have accounted for the Ni I blending transition using the Ni abundances
determined for each individual star from isolated Ni lines (see below).
Some stars artificially 
show very high values due to blending with telluric lines (above the dashed
line).
% and have been excluded from the Figure. 
HIP 109176 is 
an exception, but it is a spectroscopic binary and our spectra show
signs of a second star. 
HIP 4148 shows an unusually low upper limit, but the
[O I]- Ni I blend appears affected by  emission. 
The rest of the stars appear aligned in a pattern very similar to that found
by Nissen et al. (2002) for $-0.7 < $ [Fe/H] $< +0.25$, with a slope of
$\simeq -0.5$ below [Fe/H] $\sim 0$ that flattens out 
at super-solar metallicities.

The {\bf Magnesium} abundance is derived from the Mg I lines at
473.0, 631.9 and 880.6 nm. The strongest line at 880.6 gives a lower abundance
than the other two by 0.2-0.4 dex, but it is unclear which is to be
preferred. 
Our differential analysis, however, largely suppresses such systematic 
differences. This element shows a pattern similar to other alpha
elements, but with somewhat more `noise'. 
The alpha elements {\bf Si} and {\bf Ti}
show similar trends. {\bf Sc} mimics their behavior (compare in 
particular with Ti). The differences in the abundances of these
elements among stars at a given [Fe/H]
are very likely real, as they appear the same in the three panels. 
Silicon is measured through numerous lines of neutral Si that show quite 
consistent abundances for solar and metal-poor stars, but their behavior
degrades slightly for [Fe/H]$>0$, even after the abundances derived from
two lines in our list that are 
likely afflicted by blends are rejected. The  abundances from 
different Sc and Ti lines, however, are internally 
consistent in our full range of [Fe/H]. As these two
elements are fully ionized in the entire temperature range of our stars,
and we use ionized lines, the derived abundances 
are extremely robust to errors in the stellar parameters.
% This statement above is particularly true for Sc, perhaps due to the serious
% hfs splitting? changes in Teff by more than 300 K induced changes in the
% abundances < 0.01 dex

For metal-poor stars, the ratio 
of Si, Ti, or Sc to iron grows 
steadily from solar reaching $\sim 0.4-0.5$ dex at
[Fe/H] $\sim -1$, suggesting a slope between $-0.3$ and  $-0.4$ dex per dex
-- significantly steeper than in some of the previous studies. 
At super-solar metallicities, the trend reverses. This feature contrasts with
other analyses, which found such ratios approximately flat in  
metal-rich disk stars (e.g. Bodaghee et al. 2003; 
Feltzing et al. 2003; Chen et al. 2003; 
but see Feltzing \& Gonzalez 2001,  who
reported an average [Si/Fe] $=+0.12$ for seven metal-rich stars and
 [Ti/Fe] as high as $+0.38$ for HD 32147).

The stars with a galactic rotational  velocity less than about 
$-62$ km s$^{-1}$
(or $-50$ km s$^{-1}$ relative to the local standard of rest), shown
in red in Figs. \ref{xfe}-\ref{xfe3},  tend
to have abundance ratios near the upper envelope in the 
trends observed for Mg, Si, Ca, Sc, Ti, Co, Cu, Zn, and Eu, 
in agreement with previous analyses (e.g., 
Reddy et al. 2003; Feltzing et al. 2003) but close to
the lower envelope for Ba, Ce, and probably Nd.

{\bf Calcium} was studied with the wings of the strong Ca II lines at 849.8
and 866.2 nm, obtaining fairly similar abundances from both features. The
final abundances, however, were derived only from the 866.2 nm line, as 
this was less contaminated by blending features in the most metallic spectra.
This element's ratio to iron appears flattened compared to Si, Sc or Ti.
{\bf Nickel} shows a remarkably flat ratio to iron  for
metal-poor stars, in agreement with the results from F-type stars 
(e.g. Reddy et al. 2003). The pattern, however, changes abruptly 
for the most metal-rich stars.  At the coolest
temperatures for our stars, a variation of $\sim$ 300 K can alter the
abundances of our two Ni I lines by 0.05 dex, although in opposite senses
due to their different excitation energy. 
Ni has an ionization potential of 7.64 eV,
which is close to the 7.90 eV for iron. The small difference in
energy might be enough to introduce significant 
systematic differences in the variation of
their ionization  balances, and therefore
little weight should be given to the high abundances found for the most
metal-rich (and also the coolest) stars in our sample.
{\bf Cobalt}, however, is ionized with 7.88 eV, and therefore the ratio
Co/Fe derived from our 
list of Fe I and Co~I transitions should be a somewhat more  
robust quantity. This is particularly true for the coolest stars in the
sample, for which Co I and Fe I are the dominant species. We observe
that [Co/Fe] increases for lower [Fe/H] values, but the most metal-rich
stars show even large enrichments in cobalt.

{\bf Copper} is studied with the Cu I lines at 510.55, 521.82 and 578.2 nm.
The abundances they provide are in fairly good agreement for solar-like
metallicities and more metal-poor stars, but 
progressively diverge for higher 
metallicities\footnote{The analysis of the line
profiles in the sky spectrum with a MARCS model leads to 
$\log \epsilon {\rm (Cu)} = 4.18$ dex, in good agreement with the
 value proposed by Grevesse \& Sauval (1998) of 4.21, but slightly higher
 than derived by Cunha et al. (2002) and Simmerer et al. (2003) (4.06 dex).}.
The flat trend with increased dispersion close to solar metallicity
is not new (see e.g. Reddy et al. 2003), but the enrichment for 
metal-rich stars has not been reported before. We have
only used the line at 578.21 nm, which gives lower abundances than other
transitions in the most metal-rich stars.
This feature, again, is
insensitive to large corrections in $T_{\rm eff}$. Nevertheless, it would be 
very useful to carry out a more detailed study checking for possible blends
in the high-metallicity regime. If real, the large enhancement in Cu/Fe 
is opposite to the values found in very metal-poor halo stars in the
field or clusters (Mishenina et al. 2002; Simmerer et al. 2003).

We analyzed the Zn I transitions at 472.22 and 481.05~nm, which provide
similar abundances for metal-poor stars, but growing discrepancies at larger
metallicities. We noticed that the line at 472.22 nm has significant damping 
wings, which are systematically underestimated by the \"Unsold approximation
(adopted lacking better data), leading to higher abundances for metal-rich
stars. Thus, we decided to adopt the abundance derived exclusively from 
the line centered at 481.05 nm.
 {\bf Zinc} appears more abundant than iron in metal-poor stars than 
at solar metallicity, and even more enhanced in super-solar 
metallicity stars. 

{\bf Yttrium} shows a ratio to iron roughly flat, perhaps with a slightly
negative slope.
The {\bf Ba} abundances, as derived from the resonance line at 455.4 nm and the
high-excitation transition at 585.4 nm, show significant scatter, in 
agreement with previous results (Reddy et al. 2003). 
The other neutron-capture elements in Fig. \ref{xfe3}, 
{\bf Ce}, {\bf Nd}, and {\bf Eu},  
show less scatter as their s-process contribution decreases, 
although caution should be exercised, 
as the abundances of some of these elements are derived 
from a single feature.
 
Despite the scatter, our {\bf Eu} abundances
 indicate a  decrease in the ratio Eu/Fe with metallicity just slightly faster
 (slope of $-0.5$) than
  proposed by Woolf et al. (1995; slope $\simeq -0.4$) or Koch \& Edvardsson 
  (2002; slope $\simeq -0.3$), and  especially
   Reddy et al. (2003; slope $\sim 0$). The differences with Woolf et al.
   and Kock \& Edvardsson could be related to
   the Fe scale, which in their case was set by Fe II lines (the dominant
   species for the temperatures of their stars), or  to sampling
   of different populations within the disk.
 In our favor we should stress that 
 our analysis of this element benefits from the recent laboratory study of
 Eu II by Lawler et al. (2001)
 not available to most authors above. A satisfying agreement (within 0.15 dex)
 between the 
 abundances that we derive from the resonance line at 412.97 nm and the
 excited line at 664.51 nm supports our confidence in the new values.
We have only used the abundances from the redder feature,
which is weaker but we believe cleaner. 
Kock \& Edvardsson (2002) also found a non-zero Eu/Fe ratio at solar
metallicity ([Eu/Fe] $= +0.04$), albeit slightly lower than 
ours ([Eu/Fe] $= +0.15$). 

\begin{figure*}[t!]
\centering
{\includegraphics[width=10.2cm,angle=90]{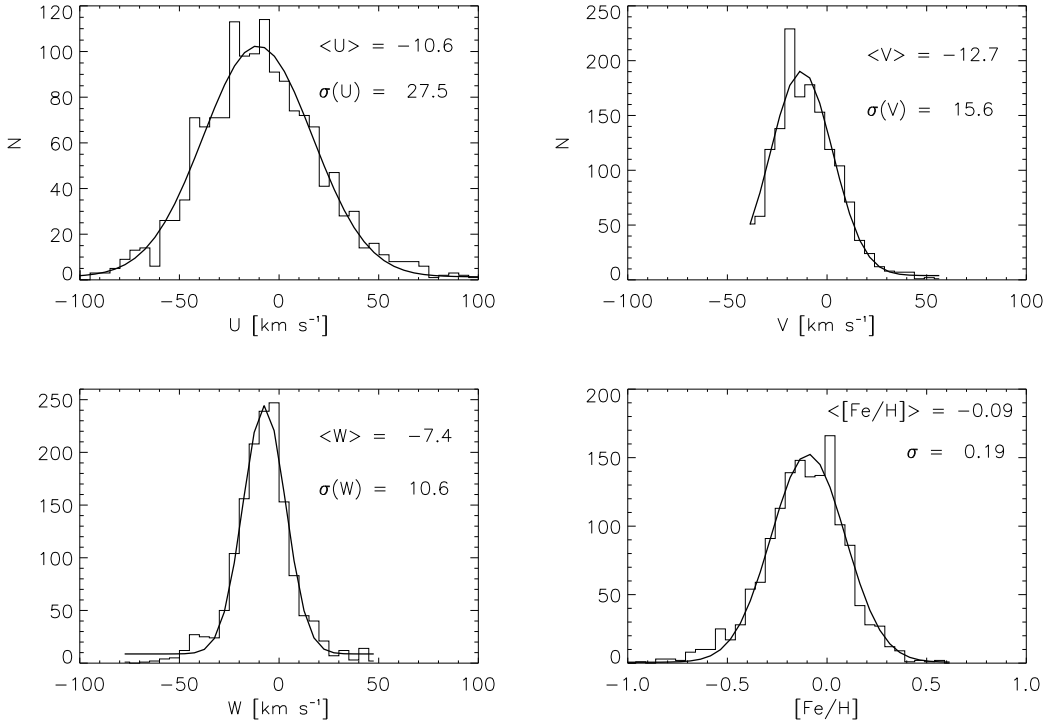}}
\caption{Velocity and metallicity histograms of the stars in our extended
sample 
within the region V$> -40$ and 
$-80<{\rm W}<+50$ km s$^{-1}$. The expected  contamination 
by thick disk and halo 
populations is very small, and therefore we identify the 
observed distributions with the thin disk.}
\label{thinb}
\end{figure*}

\section{Representativity of the sample}
\label{extended}

In this study we offer for the first time a census of nearby stars limited
in absolute magnitude and distance. Nonetheless, 
by limiting the distance to only 14.5 pc
 we are sampling a volume so small that peculiarities of some stars
could be erroneously interpreted as representative of the thin disk. 
Due to {\it Hipparcos'} completeness limits, expanding significantly
the volume 
for a complete sample requires restricting the analysis to earlier spectral
types. We can, however, compose extended samples for which some properties
are readily available and  explore whether some of the characteristics
derived from the S$^4$N stars are more generally shared by stars in the local 
thin disk. Unfortunately, at this time an expanded sample requires use of
 different, incomplete, catalogs, inevitably introducing new and complex
 selection effects.  Keeping this
 in mind, we proceed to carry out such an exercise. 

\subsection{Kinematics, metallicity, and age for an extended sample}

We collected a sample of stars whose 
astrometric positions, proper motions, and parallaxes 
were included in the {\it Hipparcos} catalog and whose
 radial velocities were part of 
 the {\it Hipparcos} input catalog, or those
compiled by Barbier-Brossat \& Figon (2000) and  Malaroda, Levato \& Galliani
(2001). Using a 0.05 degree window to match stars in the 
{\it Hipparcos} catalog
with the Barbier-Brossat \& Figon radial velocity catalog we 
identify 23742 
stars in common. We found 9312 stars in common between  
{\it Hipparcos} and the Malaroda et al. catalog, 
which concentrates on fainter stars than
Barbier-Brossat \& Figon's.  With a smaller 0.01 deg 
window we uniquely match 19072 stars 
with radial velocities in the {\it Hipparcos} input catalog 
with their counterparts in {\it Hipparcos}. A total of 28416 stars were 
in one or more of the radial velocity catalogs and {\it Hipparcos}.  
We searched for metallicities ([Fe/H]) for the sample stars 
in the catalog of  Cayrel de Strobel, Soubiran \& Ralite (2001).   
This catalog includes metallicities collected from the literature 
based on one or more sources for 2427 field stars and
582 stars in clusters and associations. Our `extended' 
sample of stars with metallicities, astrometry, and radial velocities
includes 2144 objects.

If we restrict the analysis to the region with V $> -40$ km s$^{-1}$
and $-80<{\rm W}<+50$ km s$^{-1}$, 
contamination by the thick disk and halo should be negligible. Fig. \ref{thinb}
reveals that the
distribution of stellar velocities can be reasonably well modeled by 
Gaussians in each direction. 
Kinematically, these stars belong to the same
population as those in the S$^4$N sample. The slight differences in the
widths for each velocity component can be easily explained by the statistical
errors in the smaller sample and the different distribution of spectral types
(this affects mostly V; both samples are dominated by late-type stars, but
with somewhat different proportions). 
The zero offsets -- the solar peculiar motion with
respect to its neighbors -- derived from the extended sample agree with those 
determined from the S$^4$N stars within 1 km s$^{-1}$. 
Some spikes in the velocity distributions shown in 
Fig. \ref{thinb} are statistically significant. These are well-known 
real structures in the solar neighborhood. Although the observed star counts 
 are modeled here as a single Gaussian 
 distribution, the velocity structure of the 
 solar neighborhood is quite complex. Chereul, Cr\'eze \& Bienaym\'e (1999)
 have studied the positions and kinematics of 
 A-F  {\it Hipparcos} stars within 125 pc from the Sun, identifying several
 well-known groups 
 as well as some new clusters, associations and moving groups. For 
 example, in a plot of W vs. U or V vs. U, the Hyades cluster
 can be immediately recognized as a clump centered at (U,V,W) $\simeq 
 (-43,-18,-2)$ km s$^{-1}$, which is clearly visible in the panel for V
  in Fig. \ref{thinb}. Far from being in perfectly circular orbits,
 the thin disk stars in the solar neighborhood -- and likely elsewhere-- show
 a rich structure in phase space (see, e.g., Eggen 1998).

Fig. \ref{thinb} also shows the metallicity distribution for thin disk stars.
 It is centered at [Fe/H]$=-0.09$ and has a 1-$\sigma$ width of 0.19 dex. 
 This width is 
 obviously convolved with the measurement errors. 
 In this particular case, given the very inhomogeneous
character of the Cayrel de Strobel et al. catalog, offsets between the 
metallicity scales of different authors are expected to 
contribute significantly to the observed width, yet, it is virtually
identical to the estimates from the S$^4$N sample.

\begin{figure}
\centering
{\includegraphics[width=8.5cm,angle=0]{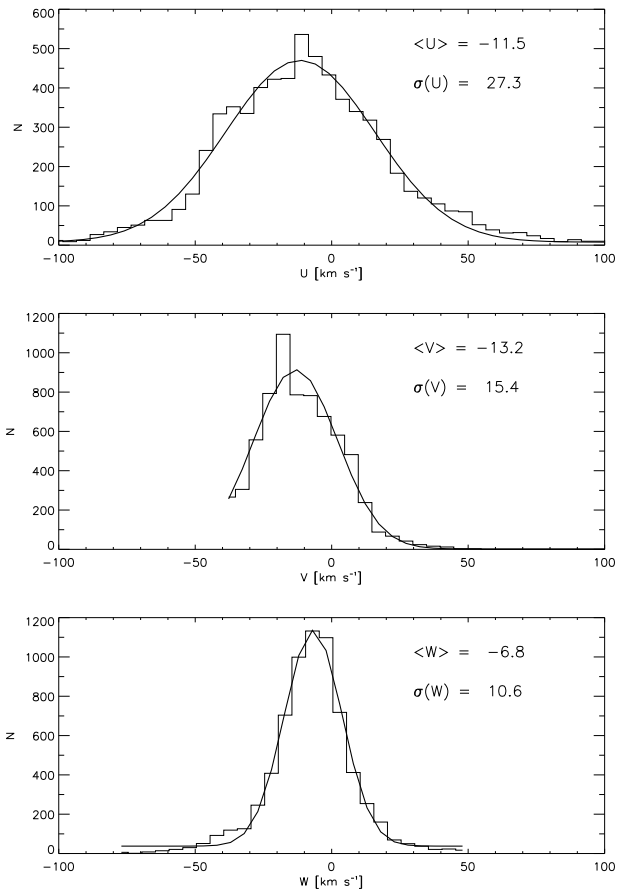}}
\caption{Velocity histograms of the stars  within 100 pc from the Sun 
in the extended sample for which 
astrometry and radial velocities, but not necessarily 
spectroscopic metallicities, 
are available. The sample is further restricted  to V$> -40$ and 
$-80<{\rm W}<+50$ km s$^{-1}$ to exclude halo and thick disk stars.
}
\label{f100pc}
\end{figure}

We can repeat the fit to velocities of the thin disk stars in our extended
sample for which we have kinematics, but not spectroscopic metallicities.
 The results for those stars within 100 pc -- a restriction imposed 
 in order to contain the errors in the astrometry-- are in good
agreement with those derived from the more restricted sample, as
shown in Fig. \ref{f100pc}. 
The Gaussian distributions become a poorer model 
for the observed distributions.

We can conclude that our sample of stars within 14.5 pc is representative,
kinematically and chemically, of a much 
larger volume of the local thin disk.

\subsection{On the peculiarity of the solar abundance ratios}

The distributions of the Si, Sc, and Ti abundance ratios relative to iron 
at solar metallicity show a positive offset with respect to the solar ratios. 
Even if we restrict the comparison to stars with atmospheric 
parameters close to solar, this offset remains.
It is difficult to attribute this to systematic errors.
Our sky spectrum, obtained with the same instrumentation and setup as the 
stellar spectra, is in very good agreement with 
a smoothed version (to compensate for the higher spectral resolution)  of 
the solar flux atlas of Kurucz et al. (1984). 
The solar analysis is identical to the stellar analysis, except that 
the atmospheric parameters for the Sun were not derived in the same manner as for
the stars.  However, this is unlikely to induce such a shift in the abundances of
different elements. 
Therefore we are inclined to suggest that the Sun is probably deficient in
these elements compared to its immediate neighbors. 
This situation may also apply to O, Ca, Y, La, Ce and Nd. 

 Edvardsson et al. (1993) found a similar effect but when they restricted the
 comparison to stars with the same ``birth galactocentric distance'' and age
 as the Sun the differences turned out to be
 insignificant. 
 Most stars in our sample, however, have
 solar-like (or thin-disk like) kinematics, so if their orbits are integrated
 most of them will give similar  galactocentric  distances at birth.
 A second issue is age. The Sun has an age in the middle of the age 
 distribution for the stars in Edvardsson et al. sample, and the density of
 stars was roughly constant at all ages, but the sample is
 afflicted with severe selection effects. 
 
\begin{figure}
\centering
{\includegraphics[width=6.cm,angle=90]{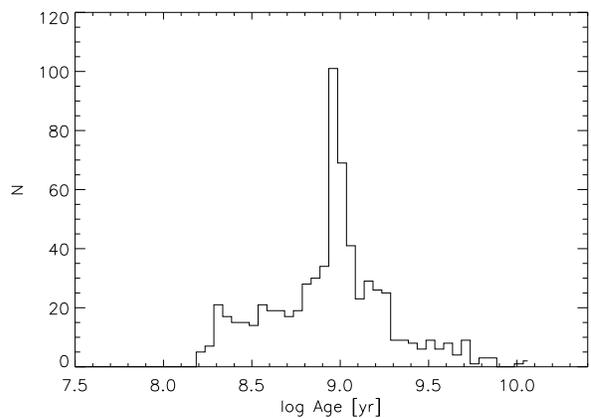}}
\caption{Age histogram of the stars in the extended sample 
 within the region 
V$> -40$ and 
$-80<{\rm W}<+50$ km s$^{-1}$ with 2-$\sigma$ uncertainties 
smaller than 0.5 dex.}
\label{thinbage}
\end{figure}

 Most of the stars in our very local (S$^4$N) sample are still relatively 
 close to the zero age main sequence (ZAMS) and therefore, we cannot derive
 ages for them by the methods described in \S \ref{others}. 
 Fig. \ref{thinbage} displays a histogram of the ages derived 
for the stars for which a 2-$\sigma$ precision of 0.5
dex or better could be achieved in our extended sample (see \S \ref{extended}). 
The age distribution covers
from  0.16 Gyr ($\log$ Age $\sim 8.2$) to 10 Gyr, with a major concentration 
at $\sim 1$ Gyr. This result is in qualitative agreement with the 
results of Chereul et al. (1999) and Feltzing et al. (2001). 
Interpretation of this distribution is a major task and 
falls outside the scope of this paper. 
We refer the reader to the recent papers
by  Bertelli \& Nasi (2001), Binney, Dehnen \& Bertelli (2002), and
Vergely et al. (2002). The key fact that we will highlight is that the Sun
($\log $ Age $\sim 9.7$)
is among  the oldest stars in our extended sample. 

The S$^4$N sample, we have seen, 
 includes a number of stars which are older than the Sun. Being 
 dominated by K-type dwarfs, it is natural old stars will
  appear preferentially in Fig. \ref{correla1}. K-type stars tend to be
 excluded from abundance studies, which prefer cleaner and brighter targets.
 Thus, metal-rich K-type dwarfs which are ubiquitous in S$^4$N are 
 discriminated against in the extended sample.
 Selection effects, however, for G-type stars are most certainly less serious.
 Therefore,  it is plausible that the G-type stars in S$^4$N 
 share the age distribution of the extended sample. Then, if the Sun is
 somewhat older than the other G-type dwarfs in S$^4$N, this would
 provide an explanation for the low abundance ratios to iron of some elements.
 The peculiarity of the Sun would be limited to its age: older
 than most thin disk dwarfs.

\section{Summary and Conclusions}
\label{wrapping}

Our spectroscopic analysis of the closest stellar neighbors reveals several 
interesting features. 
The derived [Fe/H] distribution for our sample
  is centered at [Fe/H] $\simeq -0.1$, a value 
  slightly  higher  than in most previous studies, 
  which concentrated mainly on earlier spectral types.
For a subsample of stars, comparison with evolutionary
models can provide age estimates with moderate or small uncertainties. 
This reveals a number of metal-rich stars that appear to be as old as the
most metal-poor stars in the sample ($\ga 10$ Gyr), 
and older than most solar-metallicity
thin-disk stars. These old metal-rich stars show thin-disk kinematics, and
disrupt a mild correlation between age and metallicity suggested by
the stars with [Fe/H]$<0$.

We find that our LTE abundance analysis does not
satisfy the ionization balance for several elements. The discrepancies
grow with metallicity, and appear to decrease with effective temperature, 
reaching extreme levels (nearly 0.5 dex) for
the coolest and most metal-rich stars. Similar discrepancies were previously 
reported
by Feltzing \& Gustafsson (1998) for several K-type metal-rich field dwarfs
as well as Schuler et al. (2003) and Yong et al. (2004)
 for K-type members of the open clusters M34 and the Hyades, respectively.
Systematic errors in the trigonometric gravities can be ruled out, as
satisfying the ionization balance by modifying the gravity would require 
lower values by as much as 0.8 dex. 

Systematic errors in the $T_{\rm eff}$
appear unlikely, as they would need upward revisions of up to $\sim 300$ K.
We should emphasize that 
i) the sample of stars used by Alonso et al. (1996) 
to build the photometric calibrations employed here 
reach  metallicities as high
 as [Fe/H]$=+0.5$ dex, covering adequately the range of (Fe I) 
abundances of our stars,
ii)    there is good agreement between the 
$T_{\rm eff}$s derived from Str\"omgrem photometry and the $(B-V)$
color index, and 
iii)  the temperatures derived
from the IRFM calibrations and from fitting the wings of Balmer lines are
also in good agreement. 
The only requirement of
the IRFM from model atmospheres is to correctly predict the IR continuum flux.
The fact that fluxes for the stars used in the Alonso et al. calibrations 
were obtained with the Kurucz (1993) model atmospheres, 
which have been tested here and show 
%as well as the MARCS models 
the ionization imbalance, indicates an inconsistency between observations and
the predictions of such models.
We recall that 
adopting other families of LTE line-blanketed homogeneous model atmospheres
is of no help. 
The problem is probably connected to some of  the
assumptions involved in the modeling (maybe the {\it usual suspects}: 
LTE  and surface convection). 
A direct comparison, however, of high-accuracy spectrophotometry with
model fluxes should be carried out, before it is possible to definitely state
that the inconsistencies are not related to uncertainties 
in the $T_{\rm eff}$ scale. 
Further study of open clusters, double, and multiple systems
will help to clarify the source of the  problem described.

Despite the the ionization problem, we use spectral lines of  
species that are dominant in K-type atmospheres to investigate abundance ratios
to iron for a number of chemical 
elements. Our study suggests that the $\alpha$ elements and
Sc are enhanced relative to iron for metal-rich stars. 
In an analysis strictly differential to the Sun, 
we find that some elements' ratios to iron are higher than 
solar at [Fe/H] $= 0$.
This effect is detected, for instance, 
for most  $\alpha$ elements and amounts roughly to $0.1$ dex. This fact 
could be connected to, if not explained by, the Sun being somewhat older than 
most other dwarfs with similar spectral types in the thin disk -- a conclusion
that we reach from examining an extended sample of {\it Hipparcos} stars with
metallicities from the literature.

Our efforts to obtain high-quality spectra for 
a distance- and $M_V$-limited sample of stars, have resulted 
in the identification 
of some serious problems in classical abundance analyses, rather than 
the very accurate relative abundances we hoped for. The spectra used
in this work have been placed in a public archive. 
We propose this dataset as a
standard test for modeling techniques to pass before the retrieved
abundances are to be fully trusted.

\begin{acknowledgements}

We are grateful to the staff at McDonald and La Silla Observatories, in
particular David Doss, for their helpful and professional assistance.
Comments from  the referee, Thierry Forveille, and 
from Lennart Lindegren helped to improve contents and presentation. 
We have made extensive use of the {\it Hipparcos} catalog, NASA's ADS,
and the SIMBAD databases. We gratefully acknowledge funds for this
project from the NSF (grant AST-0086321), the 
Robert A. Welch Foundation of Houston (Texas), and 
the Swedish Research Council.
\end{acknowledgements}

\section{Appendix A: Determination of stellar ages and surface gravities}

We estimated ages for the  stars in the sample by comparing the position 
in the $M_V-T_{\rm eff}$ plane of the star with the isochrones published by
Bertelli et al. (1994). Our method is similar to that described by
Reddy et al. (2003), but we have introduced some refinements. The procedure
makes no assumption on the initial mass function, the metallicity distribution,
or the star formation rate. The isochrones provide stellar properties
such as $M_V$, $T_{\rm eff}$, or $\log g$, as a function of three 
parameters: initial mass  ($M_i$),  age (Age), 
and metallicity ($Z/Z_{\odot}$). We start out by adopting  
flat distribution functions for those three parameters in the range:

\begin{equation}
\begin{tabular}{l}
$0.6 < M_i < 4.2$ \\
$6.6 < \log$ Age $< 10.2$ \\
$-1.65 < \log (Z/Z_{\odot}) < 0.35$.
\end{tabular}
\end{equation}

Iron abundances were translated to metal mass fractions assuming
$\log (Z/Z_{\odot}) =$ [Fe/H], and therefore we use them 
interchangeably below.
Adopting a Gaussian probability density for  
 $T_{\rm eff}$, $M_V$, and [Fe/H] centered at the measured values 

\begin{equation}
\begin{tabular}{cc}
$P (T_{\rm eff},M_V,{\rm [Fe/H]}) \propto$ &
$\exp \left[ -\left(\frac{T_{\rm eff}-T_{\rm eff}^{*}}{\sqrt{2} \sigma(T_{\rm eff})}\right)^2\right]$ \\
& \\
& $\times \exp \left[ -\left(\frac{M_V-M_V^{*}}{\sqrt{2} \sigma(M_V)}\right)^2\right]$ \\
& \\
& $\times \exp \left[ -\left(\frac{{\rm [Fe/H]} - {\rm [Fe/H]}^{*}}{\sqrt{2} 
\sigma({\rm [Fe/H]})}\right)^2\right]$ \\
\end{tabular}
\label{triplegauss}
\end{equation}

\noindent we determine the probability density distribution for the age

\begin{equation}
 P(\log {\rm Age}) = \int  \int P \left(T_{\rm eff},M_V,{\rm [Fe/H]}\right) 
dM_i d{\rm [Fe/H]}.
\label{edad}
\end{equation}

  In practice, to find the best age
estimate for each star, we discretized the problem by sampling the isochrones
of Bertelli et al. with constant steps of 0.006 $M_{\odot}$ 
in the initial mass $M_i$, 0.05
in $\log$ Age (Age in years), and 0.125 in [Fe/H]. 
We took advantage of a further simplification to
speed up the calculation: the Gaussian component for [Fe/H] in Eq. 
\ref{triplegauss} was replaced by a boxcar with a width of 0.25 dex after
checking that this introduced negligible differences.
We then converted
the integral in Eq. \ref{edad} into a sum over the area 
confined by an ellipsoid centered at the 
adopted values of $T_{\rm eff}$, $M_V$, and [Fe/H],
with semi-axes three times the estimated 1-$\sigma$ uncertainties in 
these parameters. 
From the probability distribution we find the mean and 
$1\sigma$ limits for the age of a star.

Similarly, the  probability density for the surface gravity is derived

\begin{eqnarray}
\begin{tabular}{ll}
 $P(\log g) =$ &  ${\displaystyle \int \int \int} \log g P [T_{\rm eff},M_V,{\rm [Fe/H]}] $ \\
\\
	      & $dM_i d{\rm [Fe/H]} d\log {\rm Age}$.	      
\end{tabular}
\label{gravedad}
\end{eqnarray}

 The isochrones employed do not consider
enhancements in the abundances of the $\alpha$ elements for metal-poor stars.
More realistic relations should take this into account, 
but as we are mainly concerned 
with thin disk stars, it will have no effect on our analysis.

\section{Appendix B: Determination of UVW velocities}
\label{uvw}

From the observed radial velocities ($V_R$), proper motions
($\mu_{\alpha}^{*} \equiv \mu_{\alpha}\cos\delta$, $\mu_{\delta}$),  
and parallaxes ($\pi$), velocities
in a cylindrical galactic system are readily obtained
(e.g., Johnson \& Soderblom 1987)

\begin{equation}
\left( \begin{tabular}{c} U \\ V \\ W \\ \end{tabular} \right)
= {\bf T A} \left( \begin{tabular}{c} $V_R$ \\ $k \mu_{\alpha}^{*}/\pi$ \\
  $k \mu_{\delta}/\pi$  \end{tabular} \right)
\label{tax}
\end{equation}

\noindent where $k=4.740470446$ is the astronomical unit in km s$^{-1}$, the 
matrix

\begin{eqnarray}
\begin{tabular}{ll}
${\bf A} = $ & 
$\left( \begin{tabular}{ccc}  $\cos\alpha$ & $\sin\alpha$ & 0 \\
$\sin\alpha$ & $- \cos\alpha$ & 0 \\ 
0 & 0 & $-1$ \\ \end{tabular} \right) $ \\
& \\
& $\left( \begin{tabular}{ccc}  $\cos\delta$ & 0 & $-sin\delta$ \\
0 &  $-1$ & 0\\ 
$-\sin\delta$ &  0 & $- \cos\delta$ \\ \end{tabular} \right)$, \\
\end{tabular} 
\end{eqnarray}

\noindent  and 
the matrix {\bf T} is a function of the equatorial position
of the galactic North Pole, 
and the position angle defining the zero of the galactic longitude

\begin{eqnarray}
\begin{tabular}{l}
${\bf T} =
\left( \begin{tabular}{ccc}  $\cos\theta_0$ & $\sin\theta_0$ & 0 \\
$\sin\theta_0$ & $- \cos\theta_0$ & 0 \\ 
0 & 0 & $1$ \\ \end{tabular} \right)$  \\
 \\
  $\left( \begin{tabular}{ccc}  $-\sin\delta_{NGP}$ & 0 & $cos\delta_{NGP}$ \\
0 &  $-1$ & 0\\ 
$\cos\delta_{NGP}$ &  0 & $ \sin\delta_{NGP}$ \\ 
\end{tabular} \right)$   \\
\\
$\left( \begin{tabular}{ccc}  $\cos\alpha_{NGP}$ & $\sin\alpha_{NGP}$ & 0 \\
$\sin\alpha_{NGP}$ & $-\cos\alpha_{NGP}$ & 0 \\ 
0 & 0 & $1$ \\ \end{tabular} \right)$   \\
\end{tabular}
\end{eqnarray}

\noindent As
all {\it Hipparcos} astrometry is given in the ICRS system J1991.25 (TT), we
use 
$\alpha_{\rm NGP}$ = 192.85948 deg, $\delta_{\rm NGP}$= 27.12825 deg, and
$\theta_0$ = 122.93192 deg (ESA 1991; vol. 1)\footnote{The corresponding angles
for B1950 are 
($\alpha_{\rm NGP}$,$\delta_{\rm NGP}$,$\theta_0$)=(192.25,25.4,123) deg
(Johnson \& Soderblom 1987).
For J2000, the angles  given for J1991.25 can be used with a negligible error 
(e.g. Murray 1989)}, and therefore {\bf T} becomes the 
transpose of {\bf $A_{\rm G}$} (Eq. 1.5.11) given in the {\it Hipparcos} catalog.

The {\it Hipparcos} catalog provides estimates of the variances and covariances
for the astrometric parameters. We can therefore improve Johnson \& 
Soderblom's estimates of the error bars in the galactic velocities by 
considering them. Given the input parameters

\begin{equation}
{\bf x} = ( \alpha ~~ \delta ~~\pi~~ \mu_{\alpha}^{*}~~ \mu_{\delta} ~~ V_R)^T
\end{equation}

\noindent The $6\times6$ covariance matrix is  adopted 
from the {\it Hipparcos} catalog

\begin{equation}
\begin{tabular}{cc}
$C_{ii}$ = $\sigma_i^2$ & $C_{ij}$ = $\sigma_i\sigma_j \rho_{ij}$ \\
\end{tabular}
\end{equation}

\noindent and the covariance matrix of the galactic velocities is calculated

\begin{equation}
{\bf J} {\bf C} {\bf J^{T}}
\label{newone}
\end{equation}

\noindent where {\bf J} is the Jacobian of the transformation in Eq. \ref{tax}

\begin{equation}
{\bf J} = 
\left( \begin{tabular}{cccc}  
%$\partial {\rm U}/\partial \alpha$ & $\partial {\rm U}/\partial \delta$ & 
%$\partial {\rm U}/\partial \pi$ & $\partial {\rm U}/\partial \mu_{\alpha}^{*}$ & 
%$\partial {\rm U}/\partial \mu_{\delta}$ & $\partial {\rm U}/\partial V_R$ \\
%$\partial {\rm V}/\partial \alpha$ & $\partial {\rm V}/\partial \delta$ & 
%$\partial {\rm V}/\partial \pi$ & $\partial {\rm V}/\partial \mu_{\alpha}^{*}$ & 
%$\partial {\rm V}/\partial \mu_{\delta}$ & $\partial {\rm V}/\partial V_R$ \\
%$\partial {\rm W}/\partial \alpha$ & $\partial {\rm W}/\partial \delta$ & 
%$\partial {\rm W}/\partial \pi$ & $\partial {\rm W}/\partial \mu_{\alpha}^{*}$ & 
%$\partial {\rm W}/\partial \mu_{\delta}$ & $\partial {\rm W}/\partial V_R$ \\
$\partial {\rm U}/\partial \alpha$ & $\partial {\rm U}/\partial \delta$ & 
\dots & $\partial {\rm U}/\partial V_R$ \\
$\partial {\rm V}/\partial \alpha$ & $\partial {\rm V}/\partial \delta$ & 
\dots & $\partial {\rm V}/\partial V_R$ \\
$\partial {\rm W}/\partial \alpha$ & $\partial {\rm W}/\partial \delta$ & 
\dots & $\partial {\rm W}/\partial V_R$ \\
\end{tabular} \right) 
= {\bf T} {\bf F}
\end{equation}

\noindent and

\begin{eqnarray}
\begin{tabular}{l}
$F_{11}=\sin\alpha (k \mu_{\delta}/\pi\sin\delta - V_R \cos\delta) - k\mu_{\alpha}^{*}/\pi\cos\alpha$ \\
$F_{12}= -V_R\cos\alpha\sin\delta - k\mu_{\delta}/\pi\cos\alpha\cos\delta$ \\
$F_{13}= k/\pi^2 (\mu_\alpha\sin\alpha + \mu_{\delta}\cos\alpha\sin\delta)$ \\
$F_{14}= -k/\pi\sin\alpha$  \\
$F_{15}= - k/\pi\cos\alpha\sin\delta$  \\
$F_{16}= \cos\alpha\cos\delta$  \\
$F_{21}= -\cos\alpha (k \mu_{\delta}/\pi\sin\delta - V_R \cos\delta) - k\mu_{\alpha}^{*}/\pi\sin\alpha$  \\
$F_{22}= -V_R\sin\alpha\sin\delta - k\mu_{\delta}/\pi\sin\alpha\cos\delta$  \\
$F_{23}= k/\pi^2 (-\mu_{\alpha}\cos\alpha + \mu_{\delta}\sin\alpha\sin\delta)$  \\
$F_{24}= k/\pi\cos\alpha$  \\
$F_{25}=-k/\pi\sin\alpha\sin\delta$  \\
$F_{26}=\sin\alpha\cos\delta$  \\
$F_{31}=0$ \\
$F_{32}= V_R\cos\delta - k/\pi\mu_{\delta}\sin\delta$ \\
$F_{33}=-k/\pi^2\mu_{\delta}\cos\delta$ \\
$F_{34}=0$ \\
$F_{35}=k/\pi\cos\delta$ \\
$F_{36}=\sin\delta$ \\
\end{tabular}
\end{eqnarray}

The uncertainties in $\mu_{\alpha}^{*}$ and 
$\mu_{\delta}$ provided in {\it Hipparcos} are defined regarding the
reference direction ($\alpha$, $\delta$) as fixed. 
Therefore, the derivatives of the 
velocity components with respect to the direction 
need not be considered, and the matrix elements 
$F_{11}, F_{12}, F_{21}, F_{22}, F_{31}$ and $F_{32}$ should be exactly
zero. The standard errors ($\sigma_i$) and 
correlation coefficients ($\rho_{ij}$) involving  $\alpha$ 
 can be substituted by
those for $\alpha^{*}\equiv \alpha\cos\delta$ 
(provided in the {\it Hipparcos} catalog), and 
$\sigma_{\alpha} \equiv \sigma(\alpha) \simeq \sigma(\alpha^{*})/cos\delta$.
Use of Eq. \ref{newone} instead of that given by Johnson \& Soderblom 
produces non-negligible differences in the estimated 
uncertainties in the galactic velocities, which generally become smaller
 for nearby {\it Hipparcos} stars.

\end{document}